\documentclass[lettersize,journal]{IEEEtran}
\usepackage{subcaption}
\usepackage{amsfonts}
\usepackage{algorithmic}
\usepackage{amsthm,amsmath,amssymb}
\usepackage{mathrsfs}
\usepackage{multirow}
\usepackage[ruled,vlined]{algorithm2e}
\usepackage{array}
\usepackage{xcolor}
\usepackage{stfloats}
\usepackage{url}
\usepackage{verbatim}
\usepackage{caption}
\usepackage{multirow}
\usepackage{graphicx}
\usepackage{color}
\usepackage{caption}
\usepackage{epstopdf}
\usepackage{cite}
\usepackage{bm}
\begin{document}
\setlength{\textfloatsep}{5pt}
\newtheorem{theorem}{Theorem}
\newtheorem{proposition}{Proposition}
\newenvironment{remark}{{\indent \it Remark:}}{}
\title{3D Extended Target Sensing in ISAC: Cram\'er-Rao Bound Analysis and Beamforming Design}


\makeatletter
\newcommand{\linebreakand}{%
  \end{@IEEEauthorhalign}
  \hfill\mbox{}\par
  \mbox{}\hfill\begin{@IEEEauthorhalign}
}
\makeatother
\author{Yiqiu~Wang,~\IEEEmembership{Graduate~Student~Member,~IEEE, }Meixia~Tao,~\IEEEmembership{Fellow,~IEEE, }\\Shu~Sun,~\IEEEmembership{Member,~IEEE, }and~Wei~Cao,~\IEEEmembership{Member,~IEEE}\vspace{-1em}
\thanks{Y. Wang, M. Tao, and S. Sun are with the Department of Electronic Engineering and the Cooperative Medianet Innovation Center (CMIC), Shanghai Jiao Tong University, China (e-mails: \{wyq18962080590, mxtao, shusun\}@sjtu.edu.cn).

W. Cao is with the ZTE Corporation, Shanghai, China. (e-mail: cao.wei8@zte.com.cn).

This work has been accepted in part for presentation at the IEEE Global Telecommunications Conference (GLOBECOM), 2024 \cite{Wang24GLOBECOM}. }}

\maketitle

\begin{abstract}
This paper investigates an integrated sensing and communication (ISAC) system where the sensing target is a three-dimensional (3D) extended target, for which multiple scatterers from the target surface can be resolved. We first introduce a second-order truncated Fourier series surface model for an arbitrarily-shaped 3D ET. Utilizing this model, we derive tractable Cram{\'e}r-Rao bounds (CRBs) for estimating the ET kinematic parameters, including the center range, azimuth, elevation, and orientation. These CRBs depend explicitly on the transmit covariance matrix and ET shape. Then we formulate two transmit beamforming optimization problems for the base station (BS) to simultaneously support communication with multiple users and sensing of the 3D ET. The first minimizes the sensing CRB while ensuring a minimum signal-to-interference-plus-noise ratio (SINR) for each user, and it is solved using semidefinite relaxation. The second balances minimizing the CRB and maximizing communication rates through a weight factor, and is solved via successive convex approximation. To reduce the computational complexity, we further propose ISACBeam-GNN, a novel graph neural network-based beamforming method that employs a separate-then-integrate structure, learning communication and sensing (C\&S) objectives independently before integrating them to balance C\&S trade-offs. Simulation results show that the proposed beamforming designs that account for ET shapes significantly outperform existing baselines, offering better communication-sensing performance trade-offs as well as an improved beampattern for sensing. Results also demonstrate that ISACBeam-GNN is an efficient alternative to the optimization-based methods, with remarkable adaptability and scalability.
\end{abstract}


\section{Introduction} 
\IEEEPARstart{T}{he} future 6G network is expected to support a wide range of location-aware services, such as autonomous driving, drone tracking, and digital twin applications. These emerging services demand high data rates, ultra-low latency, and reliable sensing capabilities. To meet these requirements, integrated sensing and communication (ISAC) has emerged as a key enabler of 6G networks \cite{Liu22JSAC}, pushing forward the shared use of radar and communication spectrums, the development of intertwined hardware architectures, and the design of dual-functional waveforms.

Despite the growing focus on ISAC, there exist many significant distinctions between radar and communication systems. One of the major differences lies in the physical models for communication users (CUs) and radar targets. In communication systems, the antenna of CUs generally has a small size and is assumed to be sufficiently distant from the base station (BS), such that the signals appear to originate from a single point in space. This point target (PT) model is similarly applicable for faraway radar targets. However, in scenarios such as autonomous driving, radar echoes may resolve multiple scatterers from target surface, since the target can have a large physical size and is close to the BS. Thus, the radar target should be more appropriately modeled as an extended target (ET).

To facilitate ET localization and tracking services, it is essential to first establish an appropriate model for the ET \cite{Karl17arXiv}. The simplest approach adopts multiple \textit{measurement sources}\footnote{Here \textit{measurement source} is used interchangeably with \textit{scatterer} to denote the physical point scattering signals from an ET.}, thereby eliminating the need to explicitly model the ET shape. The density and placement of measurement sources are determined by the employed radar sensing techniques. For instance, in the case of automotive radar echoes, the vehicle can be modeled with a fixed number of measurement sources located at the wheel-houses \cite{Hammarstrand12TAES}. In contrast, under LIDAR observations \cite{Scheel14FUSION}, the vehicle is better represented by an infinite number of measurement sources distributed continuously along its chassis. In these cases, sensing an ET is essentially equivalent to sensing several PTs, where the task is to estimate the positions of the measurement sources. To analyze the sensing accuracy for these sources, a commonly used metric is the variance bound, such as the Cram{\'e}r-Rao bound (CRB) \cite{Bekkerman06TSP}. Nevertheless, such a bound is highly unstable in the time-variant environment, where the number of measurement sources frequently changes. To address this issue, the authors in \cite{Liu21TSP} propose estimating the sensing response matrix of all measurement sources, and derive a fixed-dimension CRB for  the response matrix estimation.

While simple and easy to implement, the measurement source model has clear limitation in representing the turning maneuvers of ET. To address this issue, a shape model for the ET provides a more comprehensive solution, enabling the effective capture of the rotation and kinematic states of rigid objects. A common approach is to approximate the ET using basic geometric shapes, such as rectangles for vehicles \cite{Cao18Fusion}, ellipses for boats \cite{Liu24JSAEORS}, and lines for bicycles \cite{Haag18FUSION}. While widely adopted in target tracking, these models primarily capture the main body features of the ET, but often overlook finer structural details. For example, a real vehicle typically has rounded corners, which cannot be accurately represented by the sharp edges of a rectangular model. The limitations of geometric shape models have spurred the development of parametric shape models, which describes the ET contour with higher precision by leveraging deterministic or stochastic functions. For example, the boundary of an elliptical ET can be well represented using a single random matrix \cite{Koch08TAES}, while non-ellipsoidal boundaries can be modeled with multiple random matrices \cite{Lan16TAES}. For star-convex ETs, the authors in \cite{Hirscher16FUSION} propose a probabilistic Gaussian process contour model, which was further extended into a level-set random hypersurface model in \cite{Zea16TAES} to capture finer details along ET extensions. For irregularly shaped ETs, B-spline curves \cite{Kaulbersch18Fusion} and Fourier series \cite{Garcia22TSP} have been utilized for contour modeling. With a specific shape model, it is a non-trivial issue to evaluate the sensing performance for the ET states, including its position, kinematic parameters, and shape attributes. The authors in \cite{Xu09FUSION} and \cite{Zhong10SPL} derive the CRBs for tracking ground-moving ETs with rectangle and ellipse shapes, respectively, while the work in \cite{Sarıtaş16FUSION} extended above CRB analysis to arbitrarily shaped ETs with Gaussian process model.

Built upon the diverse modeling approaches for sensing an ET, the integration of communication tasks necessitates a unified ISAC transceiver design to address both functionalities. In this context, transmit beamforming design is crucial for achieving spatial alignment and multiplexing gains in communication, as well as spatial diversity and waveform shaping gains in radar sensing. Unlike communication, where data rate is the standard performance metric, sensing metrics vary significantly depending on the specific sensing tasks, such as target detection, estimation, tracking, or environment reconstruction, and the particular parameters to be estimated. These diverse requirements have invoked many novel ISAC transmit beamforming designs. Many early works on ISAC simply employ predetermined transmit beampatterns as the design criteria \cite{Kang19TAES}, where the transmit beams are directed towards certain desired directions for target estimation or detection. Under this paradigm, the authors in \cite{Hua23TVT} introduced sensing-dedicated signals to cover radar targets, aiming to minimize the mismatch between the desired sensing beampattern and the actual transmitted beampattern. A more sophisticated approach is to use mutual information (MI) as the radar performance metric, which, analogous to its communication counterpart, represents the theoretical limit of information that can be extracted about radar target. The authors in \cite{Li24TVT} investigated the beamforming design to maximize MI between the target response matrix and echo signals.

Note that transmit beampattern and MI lack a direct relation with the estimation accuracy of target parameters, making them less suitable for target estimation tasks. Alternatively, CRB is a more effective metric for characterizing the fundamental sensing limits in ISAC. The work in \cite{Liu21TSP} considered a monostatic ISAC system with multiple CUs and one ET, in which the CRB for estimating the ET response matrix is minimized through beamforming optimization under communication signal-to-interference-plus-noise ratio (SINR) constraint. Such a CRB is further extended to a reflective intelligent surface-aided multipath scenario, and utilized to characterize the Pareto boundary of communication and sensing (C\&S) performance trade-off in ISAC \cite{Song23TSP}. Nevertheless, for ET sensing, the primary interest typically lies in estimating its kinematic parameters of the ET rather than its response matrix. To date, no direct mapping exists between the CRB for ET response matrix and that for ET kinematic parameters. Consequently, the CRBs for estimating ET kinematic parameters, along with the corresponding CRB-based beamforming design, remain largely unexplored so far.

To our best knowledge, the work \cite{Wang24TWC} represents the first attempt in ISAC literature to adopt parametric shape model of ETs for CRB-based joint beamforming design. Using a truncated Fourier series (TFS) model for ET contour, the authors derived CRBs for ET kinematic parameter estimation, and further used the CRB of the center azimuth angle as the objective function for beamforming design. The TFS model can efficiently capture the symmetric ET shape with fewer parameters, enabling accurate representation. However, \cite{Wang24TWC} is limited to two-dimensional (2D) ETs, where the radio waves only propagate horizontally within the 2D plane and the signals coming from different elevation angles are ignored.

In this paper, we extend the contour model, CRB analysis, and beamforming design for ISAC with 2D ETs \cite{Wang24TWC} to a more realistic three-dimensional (3D) scenario. The main contributions of this paper are summarized as follows:

\begin{itemize}
    \item \textbf{Surface Model and CRB Derivation for 3D ET:} We propose an analytical second-order TFS surface model for arbitrarily shaped 3D ETs, characterized by the center range, direction, orientation, and TFS shape parameters. Based on such a model, we derive the CRB for estimating the ET kinematic parameters, and provide a fundamental analysis to unveil the main dependence of CRB upon the signal covariance matrix and ET shape. We also demonstrate that the CRB of a 3D ET can degrade to the CRB of a PT under certain circumstances.
    \item \textbf{CRB-based Beamforming Design:} We formulate two beamforming design problems, one for minimizing the sensing CRB (CRB-min problem) subject to minimum SINR constraint for each individual CU, and the other for minimizing sensing CRB and maximizing communication data rate simultaneously via a weight factor (weighted-ISAC-metric problem, WIM problem). Both problems are also subject to transmit power and ET-specific beam coverage constraints. To obtain high-quality solutions, we adopt the semidefinite relaxation (SDR) method for the CRB-min problem, and construct an iterating algorithm with successive convex approximation (SCA) technique for the WIM problem.
    \item \textbf{GNN-based Low Complexity Beamforming Design:} To reduce beamforming complexity, we propose a novel GNN-based beamforming network, ISACBeam-GNN, which leverages a \textit{separate-then-integrate} architecture to obtain the desired beamformers from C\&S input. Specifically, the sensing and communication modules in ISACBeam-GNN are designed to independently learn the nonlinear structures of CRB and SINR (or sum rate), capturing the essential characteristics of these distinct objectives. The extracted features are subsequently fused in a cascaded integration module, which optimally balances the trade-off between C\&S requirements to produce the desired beamformers. This architecture not only ensures modular clarity but also enhances adaptability, enabling ISACBeam-GNN to seamlessly accommodate various C\&S metrics, such as spectrum efficiency, MI, and others.
    \item Finally, we numerically analyze the diverse CRB characteristics for various radar targets. Compared with existing baselines, our proposed beamforming designs achieve superior C\&S performance trade-offs. Additionally, the proposed ISACBeam-GNN offers a low-complexity alternative method for beamforming that adapts effectively to varying numbers of CUs and ET scatterers.
\end{itemize}

The remainder of this paper is organized as follows. In Section II, we introduce the ISAC system model and the TFS surface model for ET. In Section III, we derive the CRB for ET sensing. In Section IV, we propose two ISAC beamforming design problems, along with the optimization-based algorithms. In Section IV, we introduce the low complexity ISACBeam-GNN beamforming design. The numerical results are illustrated in Section VI. Section VII concludes this paper.

\textit{Notations:} $\left[\cdot\right]^T$, $\left[\cdot\right]^H$, $\left[\cdot\right]^*$ denote, respectively, the transpose, Hermitian transpose, and conjugate of a matrix; $\mathbb{E}\left[\cdot\right]$ denotes the averaging operation; $\Re\left(\cdot\right)$ and $\Im\left(\cdot\right)$ respectively denote the real and imaginary part of a complex number; $\mathcal{CN}\left(\mathbf{0}_{m\times 1},\sigma^{2}\mathbf{I}_m\right)$ denotes the probability density of an ${m\times 1}$ circularly symmetric complex Gaussian vector with zero mean and covariance matrix $\sigma^{2}\mathbf{I}_m$; $\mathbb{R}^{m\times n}$ and $\mathbb{C}^{m\times n}$ denote a matrix with ${m\times n}$ real and complex elements, respectively; $\Delta _{{{\boldsymbol{\theta} }_{1}}}^{{\boldsymbol{\theta}_{2}}}\left[ \cdot \right]$ denotes the second derivative over ${\boldsymbol{\theta} }_{1}$ and ${\boldsymbol{\theta} }_{2}$; $\mathbf{A} \succeq 0$ denotes a semi-definite matrix $\mathbf{A}$; $\Vert\cdot\Vert$ and $|\cdot|$ denote the $l_2$ norm of a vector and the modulus of a scalar, respectively; $\mathbf{D}\left[\cdot\right]$ denotes a diagonal matrix; $\mathcal{F}^{0,0}(u,v)=\cos{u}\cos{v}$, $\mathcal{F}^{0,1}(u,v)=\cos{u}\sin{v}$, $\mathcal{F}^{1,0}(u,v)=\sin{u}\cos{v}$, and $\mathcal{F}^{1,1}(u,v)=\sin{u}\sin{v}$ denote the second-order Fourier series; $\otimes$ denotes the Kronecker product; $a|b$ denotes $a$ or $b$.

\section{System Model}
We consider a downlink ISAC system as shown in Fig. \ref{fig1}, where a multi-antenna BS communicates with $N_c$ single-antenna CUs and at the same time performs radar sensing for one 3D ET. We assume that the BS works in full-duplex sensing mode where the self-interference (SI) can be perfectly eliminated. To facilitate 3D ET sensing, the BS is equipped with a uniform planar array (UPA) within the $Oxz$ plane, with $N_t = N_{t,x}N_{t,z}$ transmit and $N_r = N_{r,x}N_{r,z}$ receive antennas.
\vspace{-20pt}
\subsection{3D Extended Target Surface Model}
Based on the parametric surface representation in \cite{Floreby98PFICPR}, we propose an analytical model to describe a 3D ET with a closed surface $\mathcal{S}$. As presented in Fig. \ref{fig1}, the BS is located at the origin of the global coordinate, and the ET is at the origin of the target local coordinate where the heading of the ET is set as the $+x^L$ axis. The ET orientation is defined as the angle $\varphi$ from $+x^G$ to $+x^L$. Each element along the ET surface can be uniquely identified by its position $(x,y,z)$ in the local coordinate as a function of local azimuth $u$ and elevation $v$, i.e., $\boldsymbol{\rho}\left(u,v\right)=\left[\rho_x\left(u,v\right),\rho_y\left(u,v\right),\rho_z\left(u,v\right)\right]^T$. A complete and closed surface of the ET is constituted by elements with $u\in\left[-\pi,\pi\right]$ and $v\in\left[-\pi/2,\pi/2\right]$. The corresponding $x$ component in the local coordinate by using the second-order TFS is given as
\begin{align}
&\rho_x\left(u,v\right) = \varrho_x^1 \cos{v} + \sum_{m=1}^{Q_2}\varrho_{x,m}^2\sin{lv} + \sum_{l=1}^{Q_1}\sum_{m=1}^{Q_2}\bigl[\varrho_{x,l,m}^{3}\nonumber\\
&\hspace{1.7cm}\mathcal{F}^{0,1}\left(lu,mv\right)+\varrho_{x,l,m}^{4}\mathcal{F}^{0,0}\left(lu,mv\right)\bigr],
\end{align}
where $Q_1$ and $Q_2$ are the numbers of TFS coefficients, and $\boldsymbol{\varrho}_x=[\varrho_x^1,\{\varrho_{x,m}^2\}_{m=1}^{Q_2},\{\varrho_{x,l,m}^{3},\varrho_{x,l,m}^{4}\}_{l=1,m=1}^{Q_1,Q_2}]^T$ is the TFS coefficient of $\rho_x$. The local $y\ (z)$ component ${\rho}_y\ ({\rho}_z)$ is defined similarly with mutually independent coefficients $\boldsymbol{\varrho}_y\ (\boldsymbol{\varrho}_z)$. The overall vector of TFS coefficients is denoted as $\boldsymbol{\varrho}=[\boldsymbol{\varrho}_x^T,\boldsymbol{\varrho}_y^T,\boldsymbol{\varrho}_z^T]^T$ which has a length of $N_{\boldsymbol{\varrho}}=3\left(1+Q_2+2Q_1 Q_2\right)$. Different from existing ET modeling methods, we can efficiently balance the ET modeling complexity and precision requirements with the TFS method by adjusting $Q_1$ and $Q_2$ \cite{Wang24TWC}.
\begin{figure}[!t]
\centering
\includegraphics[width=3.5in]{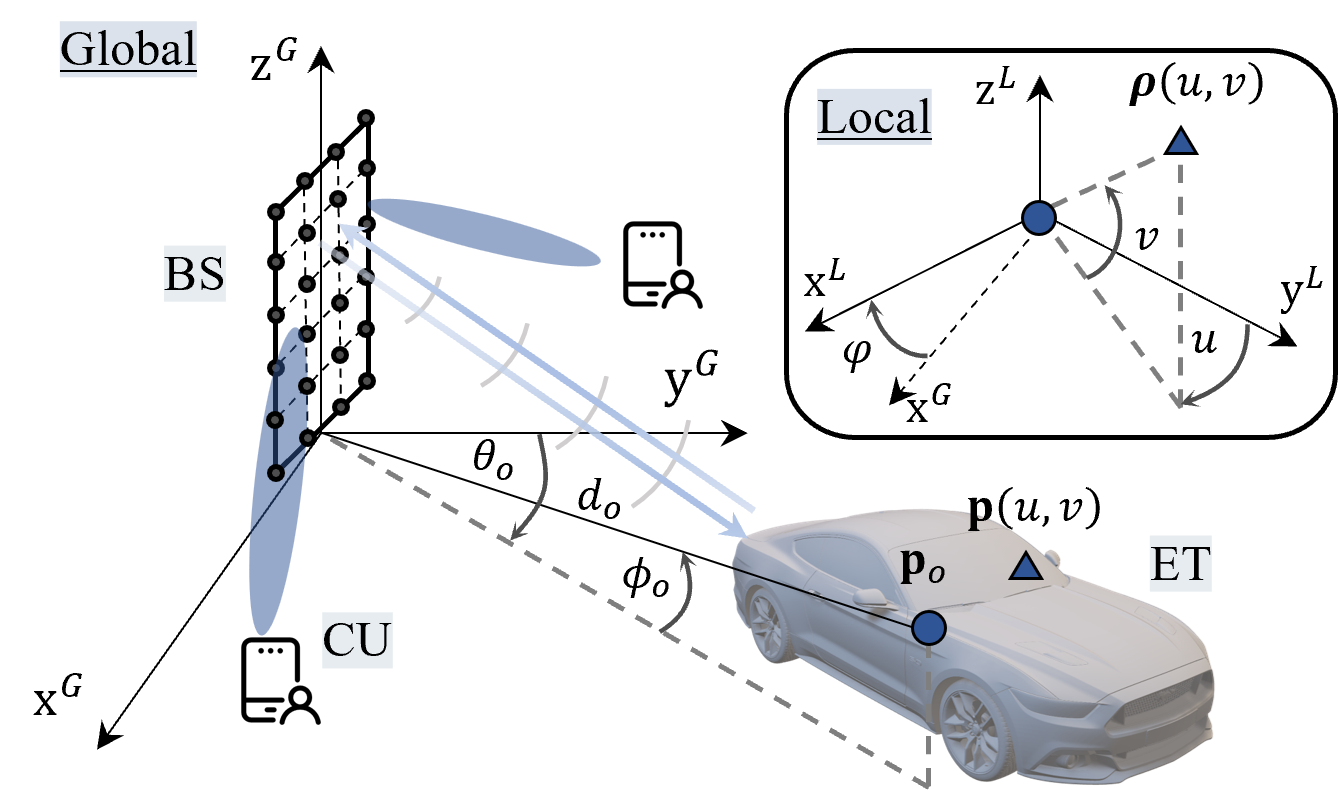}
\caption{The global and local coordinates in the ISAC system.}
\label{fig1}
\end{figure}
In the 3D global coordinate, the position of the ET center point, denoted as $\mathbf{p}_o$, is given as
\begin{equation}
\mathbf{p}_o=d_o\left[\sin\theta_o\cos\phi_o, \cos\theta_o\cos\phi_o, \sin\phi_o\right]^T,
\end{equation}
where ${d}_{o}$ is the range between the BS and the ET center, ${\theta}_{o}$ and ${\phi}_{o}$ are respectively the azimuth and elevation angles of the ET center. The global displacement of a specific element along the ET surface in local direction $(u,v)$ is expressed as
\begin{equation}
    \mathbf{p}\left(u,v\right)={{\mathbf{p}}_{o}}+\mathbf{V}\boldsymbol{\rho}\left(u,v\right),
\end{equation}
\begin{equation}
    \mathbf{V=}\left[ \begin{matrix} \cos \varphi  & -\sin \varphi & 0\\ \sin \varphi  & \cos \varphi &0 \\ 0 &0 &1\end{matrix} \right],
\end{equation}
where $\mathbf{V}$ is the rotation matrix along the $+z^G$ axis with orientation $\varphi$. The complete ET surface can thus be obtained by $\mathcal{S}=\left\{ \mathbf{p}\left( u,v \right):-\pi\le u\le \pi,-\pi/2\le v\le \pi/2 \right\}$.
\subsection{Transmit Signal Model}
In this paper, in order to maximize resource utilization, we consider the fully integrated ISAC scenario where the BS relies purely on the communication symbols for sensing without sending any dedicated radar probing signals. Let $\mathbf{x}(t) \in \mathbb{C}^ {N_t}$ denote the transmit signal from BS at time slot $t$, written as
\begin{equation}
\label{x_t}
\mathbf{x}(t) = \mathbf{W} \mathbf{s}(t),
\end{equation}
where $\mathbf{W} = \left[\mathbf{w}_1,...,\mathbf{w}_{N_c} \right] \in \mathbb{C}^{N_{t}\times N_c}$ is the overall transmit beamforming matrix for both communication and sensing, $\mathbf{s}(t) = \left[{s}_{1}(t),...,{s}_{N_c}(t) \right]^{T}$ is the information symbol intended for $N_c$ CUs. The communication symbols of different CUs are normalized and mutually independent as $\mathbb{E} \left[\mathbf{s}(t) \mathbf{s}^H (t) \right] = \mathbf{I}_{N_c}$. The covariance matrix of $\mathbf{x}(t)$ is
\begin{equation}
\label{R_x}
\mathbf{R}_x = \mathbb{E} \left[\mathbf{x}(t) \mathbf{x}^H (t) \right] = \sum\nolimits_{n=1}^{N_c} \mathbf{w}_n\mathbf{w}_n^H = \sum\nolimits_{n=1}^{N_c} \mathbf{W}_n.
\end{equation}
where $\mathbf{W}_n = \mathbf{w}_n\mathbf{w}_n^H$. The transmit signal is subject to a total power constraint as
\begin{equation}
\label{power budget}
    \mathrm{tr}\left( \mathbf{R}_{x} \right)\le {P}_{t}.
\end{equation}

\subsection{Received Communication Signal Model}
Let the communication channel from the BS to the $n$-th CU be denoted as $\mathbf{h}_n \in \mathbb{C}^{N_{t}}$, $\forall n\in\mathcal{N}$, where $\mathcal{N}=\{1,...,N_c\}$ is the set of CUs. The received signal at each CU, denoted as ${y}_n(t)$, $\forall n \in \mathcal{N}$, can be written as
\begin{equation}
\label{y_n(t)}
{y}_{n}(t) = \mathbf{h}_{n}^{H} \mathbf{x}(t) + {z}_{n}(t) = \mathbf{h}_{n}^{H} \sum\nolimits_{n=1}^{N_c} \mathbf{w}_{n}s_{n}(t) + {z}_{n}(t),
\end{equation}
where ${z}_n(t)\sim\mathcal{CN}\left(0,\sigma_n^2\right)$ is the additive white Gaussian noise (AWGN). The SINR of the $n$-th CU is expressed as
\begin{equation}
\label{SINR}
{{\gamma }_{n}}={{\mathbf{h}_{n}^{H}{{\mathbf{W}}_{n}}\mathbf{h}_{n} }}/\left({\sum\nolimits_{i=1,i\ne n}^{N_c}{{{\mathbf{h}_{n}^{H}{{\mathbf{W}}_{i}}\mathbf{h}_{n}}}+\sigma_{n}^{2}}}\right).
\end{equation}
We can further obtain the sum rate for all CUs as
\begin{align}
    R = \sum_{n=1}^{N_c}\log_2\left(1+\gamma_n\right)\label{sum rate}.
\end{align}

To facilitate correct message decoding from the received signal, a certain level of SINR threshold $\Gamma$ should be guaranteed for each CU, given as
\begin{equation}
\label{SINR constraint}
\left( 1+{{\Gamma }^{-1}} \right)\mathbf{h}_{n}^{H}{\mathbf{W}_n}{{\mathbf{h}}_{n}}\ge \mathbf{h}_{n}^{H}{{\mathbf{R}}_{x}}{{\mathbf{h}}_{n}}+\sigma _{n}^{2},\forall n\in\mathcal{N}.
\end{equation}

\subsection{Received Sensing Signal Model}

We consider monostatic sensing where the BS utilizes the received echo signal from the target for sensing. The received sensing signal at the BS is expressed as
\begin{equation}
    \label{y_s}
    \mathbf{y}_{s}\left(t\right)=\mathbf{e}\left( t \right)+\mathbf{z}_{s}\left( t \right),
\end{equation}
where $\mathbf{e}(t)\in \mathbb{C}^{{N}_{r}}$ is the echo signal, $\mathbf{z}_{s}(t)\in \mathbb{C}^{{N}_{r}}$ is the additive sensing noise following Gaussian distribution with zero mean and covariance $\sigma_s^2\mathbf{I}_{N_r}$.

For the ET, the echoes are considered as the signals scattered from the visible elements along the ET surface, whereas the energy generated from internal reflection is negligible due to the severe penetration loss. Define $\mathcal{S}_{\mathrm{visible}}$ as the visible part of the ET surface. We can divide the visible surface into $K$ nonoverlapping regions satisfying ${\mathcal{S}_{\mathrm{visible}}}=\bigcup_{k=1}^{K}{\mathcal{S}_{k}}$ and ${\mathcal{S}_{k_1}}\bigcap{\mathcal{S}_{k_2}}=\varnothing,\forall {k_1}\neq{k_2}$. For simplicity, we use \textit{scatterer} to represent the divided surface section. Consequently, the echo signal at the BS can be approximated as
\begin{align}
&\mathbf{e}\left( t \right)=\int_{\mathcal{S}_{\mathrm{visible}}}{\mathbf{e}}_{\boldsymbol{\rho}}\left( t \right)\text{d}\boldsymbol{\rho}\approx\sum\limits_{k=1}^{K}{{\mathbf{e}}_{k}}\left( t \right),\\
&{\mathbf{e}}_{k}(t)=g\sqrt{S_k}\zeta_k\mathbf{b}\left(\theta_k,\phi_k\right)\mathbf{a}^{H}\left(\theta_k,\phi_k\right)\mathbf{x}\left( t-\frac{2{{d}_{k}}}{c} \right),
\label{echo}
\end{align}
where ${{\mathbf{e}}_{\boldsymbol{\rho}}}$ and $\mathbf{e}_{k}$ refer to the echo signals as a function of $\boldsymbol{\rho}$ and $\mathcal{S}_k$, ${{\zeta }_{k}}\sim \mathcal{CN}\left(0,1 \right)$, ${\theta}_k$, ${\phi}_k$, ${{d}_{k}}$ and $S_k$ refer to the radar cross section (RCS), global azimuth angle, global elevation angle, range, and equivalent area the $k$-th scatterer, respectively, here $k\in\mathcal{K}$ and $\mathcal{K}=\{1,...,K\}$ is the set of ET scatterers, $g=\sqrt{p_0}/d_{o}^{2}\approx\sqrt{p_0}/d_{k}^{2}$ is the sensing path loss coefficient, $p_0$ is the reference path loss at $1\ \mathrm{m}$ distance, $\mathbf{a}(\cdot)$ and $\mathbf{b}(\cdot)$ are the steering vectors of transmit and receive antennas, respectively.

The transmit beampattern should illuminate the entire visible surface for reliable ET sensing. Thus, we introduce an ET-specific beam coverage requirement as
\begin{equation}
\label{beam coverage}
 \eta\min_{1 \leq k \leq K} \left( \mathbf{a}_{k}^{H}{\mathbf{R}}_{x}\mathbf{a}_{k} \right) - \max_{1 \leq k \leq K} \left( \mathbf{a}_{k}^{H}{\mathbf{R}}_{x}\mathbf{a}_{k} \right) \ge 0,
\end{equation}
where $\mathbf{a}_{k}$ is the abbreviation for $\mathbf{a}(\theta_k,\phi_k)$, $\eta\geq1$ is a pre-defined beampattern factor. The beam coverage constraint in $\left(\ref{beam coverage}\right)$ means that the max-min ratio of beam energies emitted towards every section of the visible ET surface should not exceed the given factor $\eta$. In the special case with $\eta=2$, we have a 3-dB beam coverage constraint as in \cite{Wang24TWC}.

\section{CRB Analysis for 3D Extended Target}

In this paper, we assume that the prior information of RCS parameter $\boldsymbol{\zeta }=\big[ \Re\left( {{\zeta }_{1}},{{\zeta }_{2}},...,{{\zeta }_{K}} \right)^T,\Im\left( {{\zeta }_{1}},{{\zeta }_{2}},...,{{\zeta }_{K}} \right)^T \big]^{T}\in {\mathbb{R}^{2K}}$ and the surface coefficient $\boldsymbol{\varrho}\in \mathbb{R}^{N_{\boldsymbol{\varrho}}}$ are given. Thus, the BS mainly aims to to estimate the ET kinematic parameter $\boldsymbol{\kappa }={{\left[ {d}_{o},{\theta}_{o},{\phi}_{o},\varphi\right]}^{T}}$.  Following the similar process in \cite{Wang24TWC}, the effective Fisher information matrix (EFIM) $\mathbf{J}\left(\boldsymbol{\kappa}\right)$ and the corresponding $\mathbf{CRB}\left(\boldsymbol{\kappa}\right)$ can be obtained as
\begin{equation}
\label{CRB}
\mathbf{CRB}(\boldsymbol{\kappa})=\mathbf{J}{{\left( \boldsymbol{\kappa } \right)}^{-1}}={{\left( {{\mathbf{F}}_{\boldsymbol{\kappa }}}-\frac{1}{{{f}_{g}}}{{\mathbf{f}}_{\boldsymbol{\kappa },g}}\mathbf{f}_{\boldsymbol{\kappa },g}^{T} \right)}^{-1}},
\end{equation}
where $\mathbf{F}_{\boldsymbol{\kappa}}=-\mathbb{E}\left[ \Delta _{{{\boldsymbol{\kappa }}}}^{{{\boldsymbol{\kappa }}}}\log p\left( {{\mathbf{y}}_{s}}|\boldsymbol{\xi } \right) \right]$ refers to the FIM of $\boldsymbol{\kappa}$, $f_g=-\mathbb{E}\left[ \Delta_g^g\log p\left( {{\mathbf{y}}_{s}}|\boldsymbol{\xi } \right) \right]$ is the Fisher information scalar of path loss $g$, $\mathbf{f}_{\boldsymbol{\kappa},g}=-\mathbb{E}\left[ \Delta _g^{{{\boldsymbol{\kappa }}}}\log p\left( {{\mathbf{y}}_{s}}|\boldsymbol{\xi } \right) \right]$ is the Fisher information vector regarding the path loss $g$ and $\boldsymbol{\kappa}$. In what follows, we derive a closed-form expression of $\mathbf{J}(\boldsymbol{\kappa})$ and provide some discussions.

\textit{Proposition 1:} The EFIM on $\boldsymbol{\kappa}$ can be expressed as
\begin{align}
\label{FIM}
&\mathbf{J}({\bm{\kappa}})=\frac{2g^2N_r}{\sigma_s^2}\biggl\{\sum\limits_{k=1}^{K}\Bigr[\nu_{1,k}\boldsymbol{\mu}_{1,k}\boldsymbol{\mu}_{1,k}^T+\nu_{2,k}\boldsymbol{\mu}_{2}\boldsymbol{\mu}_{2}^T+\nu_{2,3,k}\nonumber\\
&\hspace{1.2cm}(\boldsymbol{\mu}_{2}\boldsymbol{\mu}_{3}^T+\boldsymbol{\mu}_{3}\boldsymbol{\mu}_{2}^T)+\nu_{3,k}\boldsymbol{\mu}_{3}\boldsymbol{\mu}_{3}^T\Bigr]-\nu_4\boldsymbol{\mu}_{4}\boldsymbol{\mu}_{4}^T\biggr\},\\
&\text{with}\nonumber\\
&\nu_{1,k}=\frac{(4\pi B)^2}{c^2}S_k\mathbf{a}_k^H\mathbf{R}_x\mathbf{a}_k,\\
&\nu_{2,k}=t_sS_k\left(Z^{0,0}_k\mathbf{a}_k^H\mathbf{R}_x\mathbf{a}_k+\dot{\hat{\mathbf{a}}}_{k}^H\mathbf{R}_x\dot{\hat{\mathbf{a}}}_{k}\right),\\
&\nu_{2,3,k}=t_s S_k\biggl[Z^{0,1}_k\mathbf{a}_k^H\mathbf{R}_x\mathbf{a}_k+\Re\left(\dot{\hat{\mathbf{a}}}_{k}^H\mathbf{R}_x\dot{\bar{\mathbf{a}}}_{k}\right)\biggr],\\
&\nu_{3,k}=t_s S_k\left(Z^{1,1}_k\mathbf{a}_k^H\mathbf{R}_x\mathbf{a}_k+\dot{\bar{\mathbf{a}}}_{k}^H\mathbf{R}_x\dot{\bar{\mathbf{a}}}_{k}\right),\\
&\nu_{4}=\left(\sum\limits_{k=1}^KS_k\mathbf{a}_k^H\mathbf{R}_x\mathbf{a}_k\right)^{-1},
\end{align}
\begin{align}
&\boldsymbol{\mu}_{1,k}=\begin{bmatrix}1,\boldsymbol{\rho}_k^T \mathbf{V}^T\begin{bmatrix}
\mathcal{F}^{0,0}_o&-\mathcal{F}^{1,1}_o&\mathcal{F}^{0,0}_o\\ -\mathcal{F}^{1,0}_o&-\mathcal{F}^{0,1}_o&-\mathcal{F}^{1,0}_o\\ 0&\cos{\phi_{o}}&0\end{bmatrix}\end{bmatrix}^T,\\
&\boldsymbol{\mu}_{2}=\begin{bmatrix}0,&\cos^2\phi_{o},&0,&0\end{bmatrix}^T,\\
&\boldsymbol{\mu}_{3}=\begin{bmatrix}0,&0,&\cos{\phi_{o}},&0\end{bmatrix}^T,\\
&\boldsymbol{\mu}_{4}=\biggl[0,\hspace{0.3cm}\sum_{k=1}^{K}S_k\Re\left(\mathbf{a}_k^H\mathbf{R}_x\dot{\hat{\mathbf{a}}}_{k}\right)\cos^2{\phi}_{o},\nonumber\\
&\hspace{1.2cm}\sum_{k=1}^{K}S_k\Re\left(\mathbf{a}_k^H\mathbf{R}_x\dot{\bar{\mathbf{a}}}_{k}\right)\cos{\phi}_{o},\hspace{0.3cm}0\biggr]^T,
\end{align}
where $Z^{0,0}_k=\dot{\hat{\mathbf{b}}}_{k}^{H}\dot{\hat{\mathbf{b}}}_{k}/N_r$,  $Z^{1,1}_k=\dot{\bar{\mathbf{b}}}_{k}^{H}\dot{\bar{\mathbf{b}}}_{k}/N_r$,
$Z^{0,1}_k=\dot{\hat{\mathbf{b}}}_{k}^{H}\dot{\bar{\mathbf{b}}}_{k}/N_r$, $\dot{\hat{\mathbf{b}}}_{k}=\frac{\partial {{\mathbf{b}}_{k}}}{\partial \theta_k}$, $\dot{\bar{\mathbf{b}}}_{k}=\frac{\partial {{\mathbf{b}}_{k}}}{\partial \phi_k}$, $\boldsymbol{\rho}_{k}$ and $\mathcal{F}^{0,0}_o$ are, respectively, abbreviations for $\boldsymbol{\rho}(u_k,v_k)$ and $\mathcal{F}^{0,0}(\theta_o,\phi_o)$, $B$ is the effective signal bandwidth, and $t_s$ is the observation period.

\textit{Proof:} See Appendix.{$\hfill\blacksquare$}

Note that our obtained EFIM is mathematically tractable and explicit. In particular, it depends on the transmit covariance matrix $\mathbf{R}_x$ through the beam energies $\mathbf{a}_k^H\mathbf{R}_x\mathbf{a}_k$ emitted towards the visible ET surface regions. Second, the EFIM applies to arbitrarily shaped 3D ETs, provided that the TFS surface coefficients are known in advance. The impact of shapes on the EFIM is reflected in the number of visible scatterers along the ET surface (through $K$) and their spatial distributions (through $\boldsymbol{\rho}_k$ and $S_k$). In the extreme case when the ET shrinks to a PT with $K = 1$ and $S_k = 1$, the ET EFIM in ($\mathrm{\ref{FIM}}$) degrades into the PT EFIM with an additional condition of zero elevation $\phi_o=0$. This actually corresponds to the 2D scenario where the ET is located on the same plane with the BS \cite{Wang24TWC}.

\section{Problem Formulation for Beamforming Design and Optimization-Based Solutions}
In this section, we aim to optimize the transmit beamforming vectors $\{\mathbf{w}_n\}_{n=1}^{N_c}$, by taking into account both sensing CRB and communication performance. Two different problems are formulated as detailed in the following subsections.
\subsection{CRB-Minimization Problem}
This problem aims to minimize the CRB for estimating the ET kinematic parameters, i.e., $\mathrm{tr}(\mathbf{CRB}\left(\boldsymbol{\kappa}\right))$, under the constraints of CU's SINR, transmit power budget, and beam coverage requirement. For notation simplicity, we use $\mathbf{CRB}$ to represent $\mathbf{CRB\left(\boldsymbol{\kappa}\right)}$ in the following. The optimization problem is thus formulated as
\begin{align}
(\mathcal{P}1):&\underset{\left\{ \mathbf{w}_{n} \right\}_{n=1}^{N_c}}{\mathop{\min}}\,\mathrm{tr}(\mathbf{CRB})\label{Problem P0} \\
&\mathrm{s.t.}\hspace{0.2cm}(\mathrm{\ref{power budget}}),(\mathrm{\ref{SINR constraint}}),(\mathrm{\ref{beam coverage}}).\nonumber
\end{align}

The objective function $\mathrm{tr}(\mathbf{CRB})$ involves highly non-convex matrix inversion for the EFIM $\mathbf{J}(\boldsymbol{\kappa})$ in $(\ref{FIM})$. Following \cite[Appendix C]{Lyu24ITJ}, we first introduce an auxiliary semidefinite matrix $\boldsymbol{\Omega}\in\mathbb{R}^{4\times4}$, and then the CRB-min problem $\mathcal{P}1$ is equivalent to the following problem
\begin{align}
    &\underset{\left\{ \mathbf{w}_{n} \right\}_{n=1}^{N_c},\boldsymbol{\Omega}}{\mathop{\min}}\,\mathrm{tr}\left(\boldsymbol{\Omega}^{-1}\right) \label{Problem P1}
\end{align}
\begin{align}
    &\mathrm{s.t.}\hspace{0.2cm}\left[\begin{matrix}
        \mathbf{F}_{\boldsymbol{\kappa}}-\boldsymbol{\Omega} & \mathbf{f}_{\boldsymbol{\kappa},g} \\
        \mathbf{f}_{\boldsymbol{\kappa},g}^T & f_g
    \end{matrix}\right] \succeq 0,\tag{\ref{Problem P1}{a}} \label{Problem P1a}\\
    &\hspace{0.8cm}\boldsymbol{\Omega} \succeq 0,\tag{\ref{Problem P1}{b}} \label{Problem P1b}\\
    &\hspace{0.8cm}(\mathrm{\ref{power budget}}),(\mathrm{\ref{SINR constraint}}),(\mathrm{\ref{beam coverage}}).\nonumber
\end{align}

With a constraint of $\mathrm{rank}(\mathbf{W}_{n})=1$, we can replace the beamforming vector $\mathbf{w}_n$ in $(\mathrm{\ref{Problem P1}})$ by the covariance matrix $\mathbf{W}_n$. Omitting the rank-one constraint leads to the following semidefinite programming (SDP) problem
\begin{align}
    (\mathcal{P}1.1):&\underset{\left\{ \mathbf{W}_{n} \right\}_{n=1}^{N_c},\boldsymbol{\Omega}}{\mathop{\min}}\,\mathrm{tr}\left(\boldsymbol{\Omega}^{-1}\right) \label{Problem P2} \\
    &\mathrm{s.t.}\hspace{0.2cm}\mathbf{W}_n\succeq 0, \forall n\in\mathcal{N}, \mathbf{R}_x = \sum\nolimits_{{n}=1}^{N_c}\mathbf{W}_n,\tag{\ref{Problem P2}{a}} \label{Problem P2a}\\
    &\hspace{0.7cm}(\mathrm{\ref{power budget}}),(\mathrm{\ref{SINR constraint}}),(\mathrm{\ref{beam coverage}}),(\mathrm{\ref{Problem P1a}}),(\mathrm{\ref{Problem P1b}}).\nonumber
\end{align}

The above SDP problem $(\ref{Problem P2})$ can be efficiently and optimally solved by convex optimization tools, i.e., CVX. To extract rank-one beamformers $\mathbf{w}_{n}$ from the solution $\mathbf{W}_{n}$ in $(\ref{Problem P2})$, various methods can be used, e.g., Gaussian randomization. According to \cite{Lyu24ITJ}, the total number of variables in $\mathcal{P}1.1$ is $N_c N_t^2+16$. Thus, the overall computational complexity for solving $\mathcal{P}1.1$ is $\mathcal{O}\left((N_c N_t^2)^{3.5} \log_2 (1/\epsilon)\right)$, where $\epsilon$ is the solution accuracy.

\subsection{Weighted-ISAC-Metric Problem}
In some practical scenarios, it is of interests to not only minimize the sensing CRB but also maximize the sum-rate of communication users. Moreover, the SINR requirement in CRB-min problem may become infeasible for CUs in deep fade channels. To strike a balance between sensing and communication, the second beamforming design problem is to minimize the weighted sum of the communication sum rate and sensing CRB. This WIM problem, is constrained by the transmit power budget and beam coverage requirements. We formulate it as follows
\begin{align}
(\mathcal{P}2):&\underset{\left\{ {{\mathbf{w}}_{n}} \right\}_{n=1}^{N_c}}{\mathop{\min }}\alpha \beta \mathrm{tr}(\mathbf{CRB})-(1-\alpha) R\label{Problem P3}\\
&\text{s.t. }(\mathrm{\ref{power budget}}),(\mathrm{\ref{beam coverage}}),\nonumber
\end{align}
where $\alpha\in\left[0,1\right]$ is the trade-off factor between C\&S functionalities, and $\beta$ is the factor to keep $\mathrm{tr}(\mathbf{CRB})$ and $R$ on the same order of magnitude.

The objective of problem $\mathcal{P}2$ is the combination of two non-convex functions with distinct structures. Since $\mathrm{tr}(\mathbf{CRB})$, featured with inverse matrix structure, can be well translated to a convex form in $\mathcal{P}1.1$, the basic idea to solve problem $\mathcal{P}2$ is to find a concave substitute for the fractional-structured $R$. We first rewrite $R$ as
\begin{align}
&R=R_1(\mathcal{W})-R_2(\mathcal{W}),\label{R_transform}
\end{align}
where $\mathcal{W}=\left\{\mathbf{W}_n,\forall n \in\mathcal{N}\right\}$ is the set of covariance matrices for CUs, $R_1$ and $R_2$ are, respectively, defined as
\begin{align}
&R_1(\mathcal{W}) = \sum_{n=1}^{N_c}\log_2\left(\sum_{i=1}^{N_c}\mathbf{h}_n^H\mathbf{W}_i\mathbf{h}_n+\sigma_n^2\right),\\
&R_2(\mathcal{W}) = \sum_{n=1}^{N_c}\log_2\left(\sum_{i=1,i\neq n}^{N_c}\mathbf{h}_n^H\mathbf{W}_i\mathbf{h}_n+\sigma_n^2\right).
\end{align}

It can be observed that both $R_1$ and $R_2$ are concave on the covariance matrices. Now that $R$ is expressed as the difference of two concave functions in (\ref{R_transform}), it is neither concave nor convex. Consequently, we seek for a convex substitute of $R_2$ to formulate a joint concave expression for $R$. According to \cite[Appendix A]{Qiu22TWC}, suppose that $\tilde{\mathcal{W}}=\{\tilde{\mathbf{W}}_n,\forall n\in\mathcal{N}\}$ is a given point, then concave function $R_2$ is upper bounded by its convex first-order Taylor expansion $\bar{R}_2$ near $\tilde{\mathcal{W}}$, defined as
\begin{align}
&\bar{R}_2(\mathcal{W},\tilde{\mathcal{W}}) = {R}_2(\tilde{\mathcal{W}}) + \frac{1}{\ln 2}\sum_{n=1}^{N_c}\mathrm{tr}\left[\tilde{\mathbf{D}}_n(\mathbf{W}_n-\tilde{\mathbf{W}}_n)\right],
\end{align}
where $\tilde{\mathbf{D}}_n$ is defined as
\begin{align}
&\tilde{\mathbf{D}}_n = \sum_{i=1,i\neq n}^{N_c} \mathbf{h}_i z_i^{-1}(\tilde{\mathcal{W}})\mathbf{h}_i^H,\label{Dn}\\
&z_n(\tilde{\mathcal{W}}) = \sum_{i=1,i\neq n}^{N_c} \mathbf{h}_i^H \tilde{\mathbf{W}}_i\mathbf{h}_i+\sigma_n^2.\label{zn}
\end{align}

Accordingly, $R$ is now lower bounded by its concave substitute $\bar{R}(\mathcal{W},\tilde{\mathcal{W}})=R_1(\mathcal{W})-\bar{R}_2(\mathcal{W},\tilde{\mathcal{W}})$. Ignoring the constant terms in $\bar{R}_2(\mathcal{W},\tilde{\mathcal{W}})$, we further simplify $\bar{R}$ as
\begin{align}
\hat{R}(\mathcal{W},\tilde{\mathcal{W}}) &= \bar{R}(\mathcal{W},\tilde{\mathcal{W}}) - \mathrm{constant}\nonumber\\
&= R_1(\mathcal{W}) - \frac{1}{\ln 2}\sum_{n=1}^{N_c}\mathrm{tr}\left(\tilde{\mathbf{D}}_n\mathbf{W}_n\right).
\end{align}

Substitute $\hat{R}$ for $R$ in $\mathrm{(\ref{Problem P3})}$, we obtain a joint upper bound of the objective in problem $\mathcal{P}2$, and the equivalent problem can be expressed as
\begin{align}
(\mathcal{P}2.1):&\underset{\left\{ {\mathbf{W}}_{n} \right\}_{n=1}^{N_c},\boldsymbol{\Omega}}{\mathop{\min }}\alpha \beta\mathrm{tr}\left(\boldsymbol{\Omega}^{-1}\right)-(1-\alpha) \hat{R} \label{Problem P4}\\
&\text{s.t. }(\mathrm{\ref{power budget}}),(\mathrm{\ref{beam coverage}}),(\mathrm{\ref{Problem P1a}}),(\mathrm{\ref{Problem P1b}}),(\mathrm{\ref{Problem P2a}}).\nonumber
\end{align}

Problem $\mathcal{P}2.1$ is a standard SDP problem, which can be effectively solved with convex optimization tools. For better beamforming performance, we rely on the SCA method, where the convex substitute of the original problem is solved in sequence for a stationary solution. Thus, the solution $\mathcal{W}$, obtained at the current iteration, can be updated as the given point $\tilde{\mathcal{W}}$ to start the next iteration of beamforming optimization. After several iterations, we obtain a sub-optimal solution when a smooth point is reached. The SCA based transmit beamforming algorithm for the WIM problem is summarized in Algorithm 1.

The computation complexity of Algorithm 1 mainly depends on solving problem $\mathcal{P}2.1$. Since the total number of variables in problem $\mathcal{P}2.1$ is the same with that in problem $\mathcal{P}1.1$, the computation complexity of Algorithm 1 can be calculated as $\mathcal{O}\left(N_{I}(N_c N_t^2)^{3.5}\log_2(1/\epsilon)\right)$, where $N_{I}$ is the number of SCA iterations.

\begin{algorithm}[!h]
    \caption{SCA Beamforming Algorithm for $\mathcal{P}2$}
    \label{alg:AOA}
    \renewcommand{\algorithmicrequire}{\textbf{Input:}}
    \renewcommand{\algorithmicensure}{\textbf{Output:}}
    \begin{algorithmic}[1]
        \REQUIRE $\text{CU channels}\left\{\mathbf{h}_n,\forall n\in\mathcal{N}\right\}, \text{iterating number }N_I,$ \\
        \hspace{15pt}$\text{Scatterer info}\left\{S_k,\theta_k,\phi_k,\boldsymbol{\rho}_k,\forall k\in\mathcal{K}\right\},$ \\
        \hspace{15pt}$\text{Kinematic info }\left\{\theta_o,\phi_o,d_o,\varphi \right\}, \text{shape info }\boldsymbol{\varrho}.$
        \ENSURE Beamforming vectors $\{\mathbf{w}_n,n\in\mathcal{N}\}.$   
        
    \STATE  Random initialize: $\tilde{\mathcal{W}}=\{\tilde{\mathbf{W}}_n,\forall n\in\mathcal{N}\},t=0.$
        \WHILE{$t<N_I$}
            \STATE $t = t+1.$
            \STATE Calculate $\tilde{\mathbf{D}}_n$ in $\mathrm{(\ref{Dn})}$, $z_n(\tilde{\mathcal{W}})$ in $\mathrm{(\ref{zn})}$, $\forall n\in\mathcal{N}$.
            \STATE Solve the SDP problem $\mathcal{P}2.1$ to obtain covariance matrices $\mathbf{W}_n,\forall n\in\mathcal{N}$.
            \STATE Update $\tilde{\mathbf{W}}_n=\mathbf{W}_n,\forall n \in\mathcal{N}$.
        \ENDWHILE
        \STATE Recover $\{\mathbf{w}_n,\forall n\in\mathcal{N}\}$ from $\{\mathbf{W}_n,\forall n\in\mathcal{N}\}$ via Gaussian randomization.
    \end{algorithmic}
\end{algorithm}

\begin{figure*}[!t]
\centering
\includegraphics[width=7in]{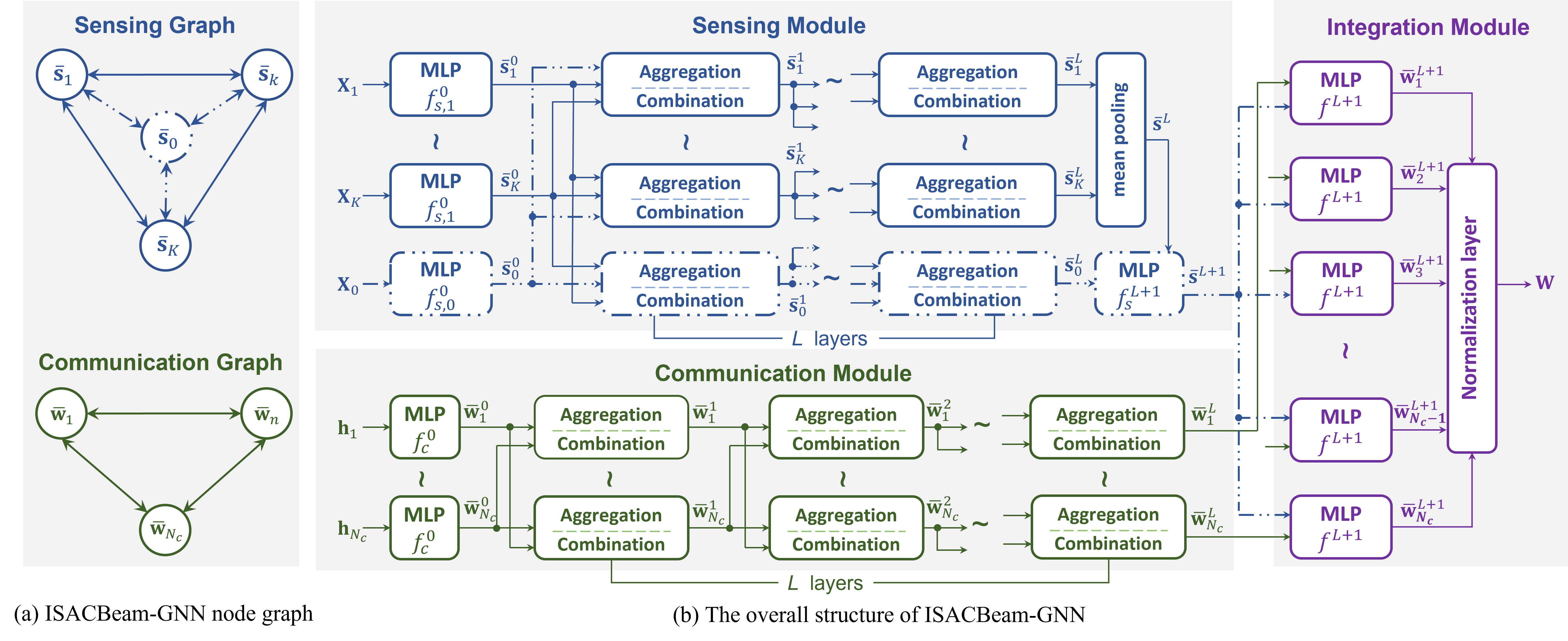}
\caption{The proposed ISACBeam-GNN.}
\label{fig2}
\end{figure*}

\section{Learning-based Beamforming Design}
Note that solving problems $\mathcal{P}1.1$ and $\mathcal{P}2.1$ suffers from high complexity of SDR and SCA algorithms, which limits their feasibility for real-time applications. In this section, we propose a low complexity learning-based approach, namely ISACBeam-GNN, to optimize the transmit beamformers for both problems. We first introduce the details of the proposed ISACBeam-GNN, and then present the learning loss functions for problems $\mathcal{P}1$ and $\mathcal{P}2$.
\vspace{-10pt}
\subsection{ISACBeam-GNN Architecture}
Since the exact numbers of CUs and ET scatterers may vary across observation periods, we employ a GNN-based framework, ISACBeam-GNN, to handle the dynamic input dimensions and optimize joint C\&S beamforming. While GNNs have been extensively used in wireless communication for physical layer design  \cite{Tao21JSAC,Zhang23TCOM}, we extend their application to ISAC domain with a novel separate-then-integrate structure, featuring three GNN modules and three types of nodes.

As shown in Fig. \ref{fig2}, ISACBeam-GNN consists of the sensing module, communication module, and integration module, which are tailored for CRB minimization, SINR satisfaction or rate maximization, and their trade-off integration, respectively. The detailed structure of ISACBeam-GNN is shown as follows

1) \textbf{Sensing Module}: It is designed to capture the nonlinear structure of the sensing CRB. The sensing node graph in Fig. $\mathrm{\ref{fig2}a}$ comprises $K$ scatterer nodes and one center node, representing the scatterer features \(\{\bar{\mathbf{s}}_k, \forall k \in \mathcal{K}\}\) and the ET center feature \(\bar{\mathbf{s}}_0\). Each scatterer node has bidirectional edges connecting to all other scatterer nodes and the center node, mimicking the scatterer-center structure in CRB formula \((\mathrm{\ref{CRB}})\).

To be specific, the sensing module takes the parameters of the scatterers $\{\mathbf{X}_k=\left[S_k,\theta_k,\phi_k,\boldsymbol{\rho}_k^T\right]^T,\forall k\in\mathcal{K}\}$ and the ET center $\mathbf{X}_0=\left[\theta_o,\phi_o,d_o,\varphi\right]^T$ as input. We use a shared multilayer perceptron (MLP) network $f^0_{s,1}(\cdot)$ to produce the feature $\bar{\mathbf{s}}_k^0$ from $\mathbf{X}_k$ for scatterer node $k$, along with a dedicated MLP $f^0_{s,0}(\cdot)$ to extract the feature $\bar{\mathbf{s}}_0^0$ from $\mathbf{X}_0$ for center node, expressed as
\begin{align}
    &\bar{\mathbf{s}}_0^0 = f^0_{s,0}(\mathbf{X}_0),\\
    &\bar{\mathbf{s}}_k^0 = f^0_{s,1}(\mathbf{X}_k),\forall k\in\mathcal{K}.
\end{align}

The update of GNN node features involves two basic functions \cite{Hamilton17NIPS}, denoted as the aggregation function $f_{ag}(\cdot)$ and the combination function $f_{cb}(\cdot)$. Before updating the node feature $\bar{\mathbf{s}}^l_k$ in the $l$-th layer, we first pre-process its previous feature $\bar{\mathbf{s}}^{l-1}_k$ and the neighbouring node features $\{\bar{\mathbf{s}}^{l-1}_j\}_{\forall j\in\mathcal{N}_k}$ with pre-process function ${f}_{pre}(\cdot)$, where $\mathcal{N}_k$ denotes the set of neighbouring nodes for the $k$-th scatterer node. Then, we combine the processed node feature with the aggregated neighboring features for node update. The update process is
\begin{align}
    \bar{\mathbf{s}}^l_k = f^l_{cb}\left(f^l_{pre}\left(\bar{\mathbf{s}}^{l-1}_k\right),f^l_{ag}\left(f^l_{pre}\left(\{\bar{\mathbf{s}}^{l-1}_j\}_{\forall j\in\mathcal{N}_k}\right)\right)\right).
\end{align}

In this work, we adopt the element-wise mean pooling operation as the aggregation function, i.e., $f_{ag}(\cdot) = \mathrm{mean}(\cdot)$, to characterize the interaction patterns between the scatterer nodes and the center node that influence the CRB value. The combination and pre-processing functions are implemented using distinct MLPs. As illustrated in Fig. \ref{fig3}, for the $k$-th scatterer node, the feature update in the $l$-th layer is

\begin{align}
    \bar{\mathbf{s}}^l_k = f^l_{s,3}\left(f^l_{s,0}(\bar{\mathbf{s}}^{l-1}_0),f^l_{s,1}(\bar{\mathbf{s}}^{l-1}_k),\mathrm{mean}\left\{f^l_{s,2}(\bar{\mathbf{s}}^{l-1}_j)\right\}_{\forall j\in\mathcal{N}_k}\right).
\end{align}

The feature update of the center node is denoted as
\begin{align}
    \bar{\mathbf{s}}^l_0 = f^l_{s,5}\left(f^l_{s,0}\left(\bar{\mathbf{s}}^{l-1}_0\right),\mathrm{mean}\left\{f^l_{s,4}\left(\bar{\mathbf{s}}^{l}_k\right)\right\}_{\forall k\in\mathcal{K}}\right).
\end{align}

Finally, after $L$ layers of updates, we distill the relevant features for CRB from the enriched feature set of all sensing nodes, expressed as
\begin{align}
    \bar{\mathbf{s}}^{L+1} = f^{L+1}_s\left(\bar{\mathbf{s}}^{L}_0,\mathrm{mean}\left\{\bar{\mathbf{s}}^{L}_k\right\}_{\forall k\in\mathcal{K}}\right).
\end{align}

\begin{figure*}[!t]
\centering
\includegraphics[width=7in]{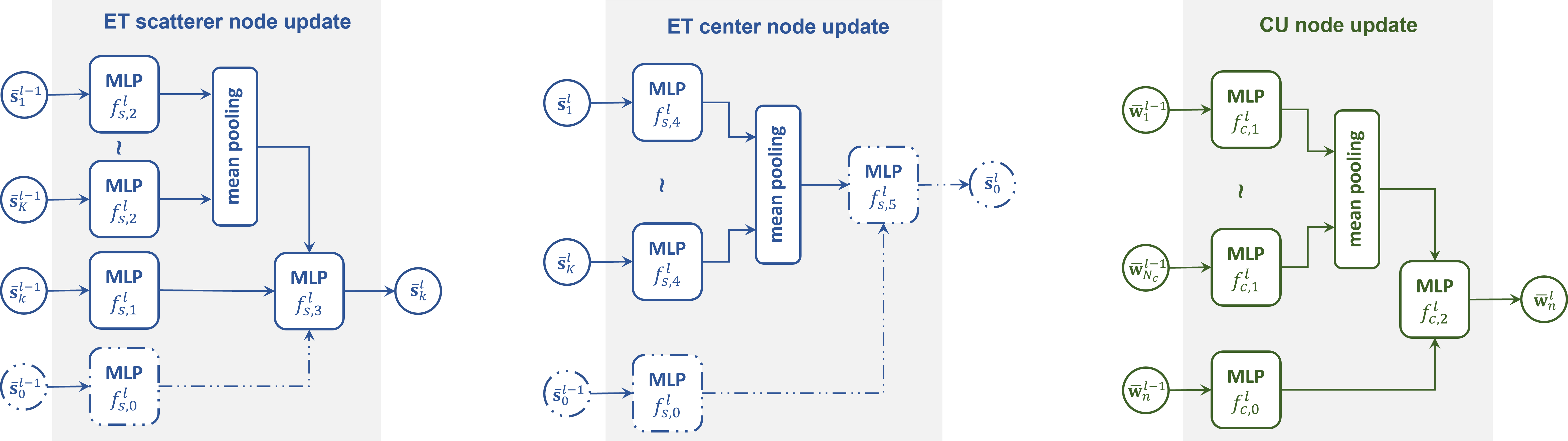}
\caption{The update of the scatterer node $\bar{s}_k^l$, the ET center node $\bar{s}_0^l$, and the CU node $\bar{w}_n^l$ in the $l$-th layer.}
\label{fig3}
\end{figure*}

2) \textbf{Communication Module}: It focuses on SINR satisfaction or sum rate maximization. As is shown in Fig. $\mathrm{\ref{fig2}a}$, the communication node graph consists of $N_c$ CU nodes, each representing the communication feature $\bar{\mathbf{w}}_n$ of the $n$-th CU. To model interference, each CU node is connected via bidirectional edges to all neighboring nodes, enabling the module to manage signal interference and separation effectively.

To be specific, the communication module has $N_c$ CU nodes, and each takes its own communication channel vector $\mathbf{h}_n$ as input to obtain the initial feature as
\begin{align}
\bar{\mathbf{w}}_n^{0}=f^0_c\left(\left[\Re(\mathbf{h}_n)^T,\Im(\mathbf{h}_n)^T\right]^T\right),\forall n\in\mathcal{N}.
\end{align}

The feature update for the $n$-th CU node in the $l$-th layer, as shown in Fig. \ref{fig3}, is given by
\begin{align}
    \bar{\mathbf{w}}^l_n = f^l_{c,2}\left(f^l_{c,0}\left(\bar{\mathbf{w}}^{l-1}_n\right),\mathrm{mean}\left\{f^l_{c,1}\left(\bar{\mathbf{w}}^{l-1}_j\right)\right\}_{\forall j\in\mathcal{N}_n}\right),
\end{align}
where $\mathcal{N}_n$ denotes the neighbouring node set of the CU node $n$. After $L$ layers of update, we obtain the refined features of SINR or sum rate from all CU nodes as $\{\bar{\mathbf{w}}^L_n,\forall n\in\mathcal{N}\}$.

3) \textbf{Integration Module}: It embodies the integrate phase of the architecture, combining refined features from sensing and communication modules to obtain the desired beamformers.

Specifically, the integration layer achieves the ISAC performance trade-off by assigning appropriate weights to the updated sensing node feature $\bar{\mathbf{s}}^{L+1}$ and communication node features $\{\bar{\mathbf{w}}^L_n, \forall n \in \mathcal{N}\}$. The integration MLP processes the combined updated features from each CU node and the sensing node, i.e., $\{\bar{\mathbf{s}}^{L+1}, \bar{\mathbf{w}}^L_n\}$, to produce the unscaled transmit beamformer $\bar{\mathbf{w}}^{L+1}_n$. To ensure compliance with the power constraint in $(\mathrm{\ref{power budget}})$, a normalization layer is appended to the integration module, scaling the beamformers accordingly. The entire process can be expressed as
\begin{align}
&\left[\Re(\bar{\mathbf{w}}_n^{L+1})^T,\Im(\bar{\mathbf{w}}_n^{L+1})^T\right]^T = f^{L+1}\left(\bar{\mathbf{s}}^{L+1},\bar{\mathbf{w}}^{L}_n\right),\\
&\mathbf{w}_n = \bar{\mathbf{w}}_n^{L+1}\times\sqrt{P_t/\left(\sum\nolimits_{i=1}^{N_c}\Vert\bar{\mathbf{w}}_i^{L+1}\Vert^2\right)},\forall n\in\mathcal{N}.
\end{align}
\begin{remark}
    The proposed ISACBeam-GNN offers several advantages in the ISAC scenario. First, it embeds the topologies for ET sensing and wireless communication directly within the node graphs, enabling effective modeling of multi-user interference in communication and scatterer-center interaction in sensing, making the whole ISACBeam-GNN interpretable. Second, by leveraging the distinct characteristics of sensing and communication functionalities, ISACBeam-GNN adopts a \textit{separate-then-integrate} approach, dividing the beamformer design into independent sensing and communication modules before final integration. This modular design provides a general learning framework for ISAC, and is highly adaptable for various metrics, such as beampattern matching error, with only minor modifications to the nodes and graph structures. Third, thanks to the uniform feature update mechanism applied to nodes of the same type, ISACBeam-GNN exhibits strong scalability and generalization. It seamlessly adapts to varying numbers of CUs and ET scatterers without requiring additional training, making it efficient for dynamic ISAC environments.
\end{remark}

The inference complexity of ISACBeam-GNN primarily depends on the computations within the MLPs. The number of flops in an MLP is $\mathcal{O}\left(\sum\nolimits_{l=1}^{N_L}n_l n_{l-1}\right)$, where $N_L$, $n_l$, and $n_0$ respectively denote the number of layers, the feature dimension of the $l$-th layer, and the input dimension, respectively. Thus, the computation complexities of the sensing, communication, and integration modules in ISACBeam-GNN are respectively $\mathcal{O}\left(2368K+15360KL+10496\right)$, $\mathcal{O}(48N_t$$N_c+117N_tN_cL)$, and $\mathcal{O}\left(8N_t^2N_c+128N_tN_c\right)$. Based on the above discussion, the overall complexity of online beamforming by ISACBeam-GNN is $\mathcal{O}\left(\max\left(KL,N_tN_cL,N_t^2N_c\right)\right)$. Compared with the complexities of solving optimization problems $\mathcal{P}1.1$ and $\mathcal{P}2.1$, ISACBeam-GNN achieves a significant reduction in flops and running time, as presented in Table \ref{tab:time}.
\begin{table}
  \caption{Average Computation Time (in seconds).}
  \centering
  \begin{scriptsize} 
  \label{tab:time}
  \begin{tabular}{|c|c|c|c|c|}
  \hline
    \multirow{2}{*}{Setting} & \multicolumn{2}{c|}{$\mathrm{CRB_{min}\ Design}$} & \multicolumn{2}{c|}{$\mathrm{WIM\ Design}$}\\
  \cline{2-5}
    & Optimization & Learning & Optimization & Learning\\
  \hline
    $K=38,N_c=4$ & 49.7813 & 0.7135 & 125.4219 & 0.7186\\
  \hline
  \end{tabular}
  \end{scriptsize} 
\end{table}
\subsection{Loss Function Design}
Since learning-based approachs are generally designed for unconstrained optimization problems, we exploit the penalty method \cite{LiuC22JSAC} to handle the constrains in problems $\mathcal{P}1$ and $\mathcal{P}2$. The proposed ISACBeam-GNN can be trained in an unsupervised manner with loss functions given by
\begin{align}
&\mathcal{L}_{\mathcal{P}1} =\rm{tr}(\mathbf{CRB})+\lambda_1\sum\nolimits_{n=1}^{N_c}\max\left(0,\Gamma-\gamma_n\right)^2+\lambda_2 \nonumber\\
&\hspace{0.6cm}\max\left[0,\max_{\forall k\in\mathcal{K}} \left( \mathbf{a}_{k}^{H}{\mathbf{R}}_{x}\mathbf{a}_{k} \right) - \eta\min_{\forall k\in\mathcal{K}} \left( \mathbf{a}_{k}^{H}{\mathbf{R}}_{x}\mathbf{a}_{k} \right)\right]^{\smash{\raisebox{-0.5ex}{$\scriptstyle 2$}}}\label{Problem P5},\\
&\mathcal{L}_{\mathcal{P}2} =\alpha \beta \mathrm{tr}(\mathbf{CRB})-(1-\alpha) R+\lambda_2\nonumber\\
&\hspace{0.6cm}\max\left[0,\max_{\forall k\in\mathcal{K}} \left( \mathbf{a}_{k}^{H}{\mathbf{R}}_{x}\mathbf{a}_{k} \right) - \eta\min_{\forall k\in\mathcal{K}} \left( \mathbf{a}_{k}^{H}{\mathbf{R}}_{x}\mathbf{a}_{k} \right)\right]^{\smash{\raisebox{-0.5ex}{$\scriptstyle 2$}}},\label{Problem P6}
\end{align}
where $\lambda_1$ and $\lambda_2$ respectively denote the penalty factors for the constraints of communication SINR threshold in 
$\mathrm{(\ref{SINR constraint})}$ and sensing beam coverage requirement in $\mathrm{(\ref{beam coverage})}$. Note that the transmit power inequality constraint in $\mathrm{(\ref{power budget})}$ is not considered here, since it can be easily implemented by attaching a normalization layer at the end of ISACBeam-GNN.

\section{Numerical Results}
In this section, we numerically analyze the derived CRBs and evaluate the proposed beamforming designs, i.e., $\mathrm{CRB_{min}\ Design\mbox{-}O|L}$ in $\mathcal{P}1$ and $\mathrm{WIM\ Design\mbox{-}O|L}$ in $\mathcal{P}2$, with a simulation study. Here $\mathrm{O}$ and $\mathrm{L}$ stand for $\mathrm{Optimization}$ and $\mathrm{Learning}$, respectively.
\vspace{-10pt}
\subsection{Simulation setup}
1) \textbf{Wireless network setting:} The BS is equipped with a UPA of $N_t = N_r = 8\times8$ elements. Unless otherwise stated, we set the transmit power constraint $P_t = 30\ \mathrm{dBm}$, the carrier frequency $f_c = 30\ \mathrm{GHz}$, the effective channel bandwidth $B = 1\ \mathrm{GHz}$, and the noise powers $\sigma_s^2 = \sigma_n^2 = -80\ \mathrm{dBm}$. There exist four CUs located at directions ($0^\circ,30^\circ$),  ($0^\circ,-30^\circ$), ($-40^\circ,30^\circ$), and ($-40^\circ,-30^\circ$). We utilize the Saleh-Valenzuela channel model for the BS-CU link with a path loss of $100\ \mathrm{dB}$. For sensing, we consider a pure line-of-sight path. The SINR threshold in $\mathcal{P}1$ is $\Gamma = 0\ \mathrm{dB}$, the trade-off and balance factors in $\mathcal{P}2$ are, respectively, $\alpha=0.5$ and $\beta=10^{4}$.

2) \textbf{3D ET setting:} The vehicle-shaped 3D ET, characterized by $Q_1 = 8$ and $Q_2 = 8$ TFS coefficients, is assumed to have a cuboid surface with a length, width, and height of $5\ \mathrm{m}$, $2\ \mathrm{m}$, and $1.2\ \mathrm{m}$, respectively. The ground truth values for the ET kinematic parameters are $d_o = 8.7\ \mathrm{m}$, $\theta_o = \varphi = 0^{\circ}$, $\phi_o = -23^{\circ}$, and we set $t_s = 1\ \mathrm{s}$, $\eta = 5$.

3) \textbf{ISACBeam-GNN setting:} We adopt a 4-layer ISAC Beam-GNN with one input layer, two hidden layers ($L=2$) and one output layer. The input and hidden layers adopts $\mathrm{ReLU}$ as the activation function, and the output layer uses $\mathrm{Linear}$ activation function. In the training phase, we choose $\mathrm{Adam}$ as the optimizer. The initial learning rate is set to be 0.001 which decreases every 10 epochs with a decay-rate of 0.96. We set the drop-out rate as 0.2 and the network weight decay as $10^{-4}$. The numbers of training, validation, and test samples, which are randomly generated based on different CU and ET locations, are respectively $10240$, $2048$ and $2048$. The training phase gets terminated if the losses of the validation set do not decrease within 50 epochs. The experiments are performed on an Intel Xeon Silver 4214R CPU, and a 24 GB Nvidia GeForce RTX 3090 Ti graphics card with Pytorch powered with CUDA 11.4. The penalty factors in $\mathcal{L}_{\mathcal{P}1}$ and $\mathcal{L}_{\mathcal{P}2}$ are set as $\lambda_1=\lambda_2=10$.

\begin{figure*}[!t]
\centering
\includegraphics[width=7in]{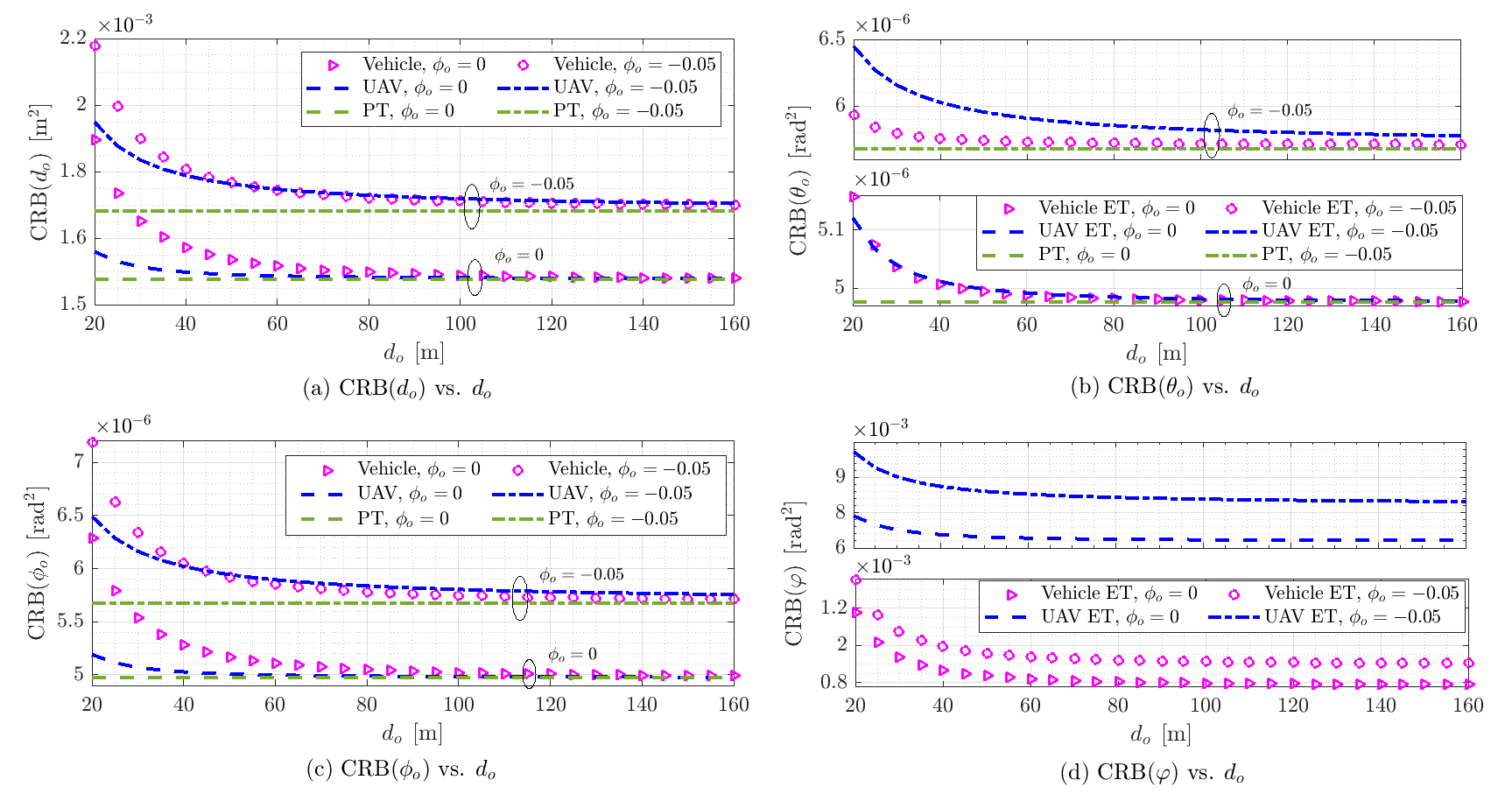}
\caption{Normalized CRBs of vehicle-shaped ET, drone-shaped ET, and PT kinematic parameters versus distance $d_o$.}
\label{fig:CRB analysis}
\end{figure*}

4) \textbf{Benchmark setting:} Two baselines are selected for comparison with our proposed beamforming designs as detailed below.
\begin{itemize}
    \item Benchmark 1 ($\mathrm{Center}$ $\mathrm{Design}$): This scheme follows the traditional beamforming design for a PT where the beam energy is maximized towards the center of the ET. Denote $\mathbf{a}_o=\mathbf{a}(\theta_o,\phi_o)$, the optimization problem is
    \begin{align}
    \underset{\left\{ {{\mathbf{W}}_{n}} \right\}_{n=1}^{N_c}}{\mathop{\min }}\mathbf{a}^H_o\mathbf{R}_x\mathbf{a}_o\nonumber
    \end{align}
    \item Benchmark 2 ($\mathrm{Average}$ $\mathrm{Design}$): This is the scheme introduced by \cite{Hua23TVT} which maximizes the minimum energy transmitted towards the given area $\mathcal{A}$ covered by the ET, without considering the specific ET shape. Denote $\mathbf{a}_i=\mathbf{a}(\theta_i,\phi_i)$, the optimization problem is
    \begin{align}
    &\underset{t,\left\{{\mathbf{W}}_{n} \right\}_{n=1}^{N_c}}{\mathop{\max }}t\hspace{0.3cm}\text{s.t. } \mathbf{a}_i^H\mathbf{R}_x\mathbf{a}_i \geq t,\forall \{\theta_i,\phi_i\} \in \mathcal{A}.\nonumber
    \end{align}
\end{itemize}

Each point in the simulation results is obtained via averaging over $2,000$ Monte Carlo realizations.

\subsection{Numerical Results on CRB}
We commence with comparing the CRBs of a vehicle-shaped ET, drone-shaped ET (arm $1.15\ \mathrm{m}$, height $0.65\ \mathrm{m}$), and a PT at different distances in Fig. \ref{fig:CRB analysis}. For the sake of analysis, the radar SNR, i.e., $P_t g^2/\sigma_s^2$, is kept fixed regardless of the distances between the targets and the BS, and the areas of the ETs are normalized to unit value, i.e., $\sum\nolimits_{k=1}^K S_k=1$. First, we observe from Fig. \ref{fig:CRB analysis}$\mathrm{a}$, \ref{fig:CRB analysis}$\mathrm{b}$, and \ref{fig:CRB analysis}$\mathrm{c}$ that, while the CRBs of the PT remain constant across all distances $d_o$, the CRBs of the ET first decrease and then converge a constant as the distance increases. Specifically, when the elevation of the ET is $\phi_o=0$ (equivalent to the 2D scenario), the ET CRBs of $d_o$, $\theta_o$, and $\phi_o$ converge to exactly the same values of the PT CRBs at large distances. But when $\phi_o \neq 0$ (which is $-0.05$ for a strict 3D scenario), there are notable gaps between the converged CRBs of ETs and those of PT. These observations are consistent with the discussion in Section III that the equivalence between ET and PT is only valid under the condition that $\phi_o = 0$. It is also interesting to discover that, for the same ET shape with the same distance, CRBs of ET at $\phi_o = 0$ are smaller than those at $\phi_o = -0.05$. This indicates higher estimation accuracy can be obtained when the elevation of the ET is 0. Moreover, from Fig. \ref{fig:CRB analysis}$\mathrm{d}$, we observe that $CRB(\varphi)$ is on the order of $10^{-3}$, which is three orders of magnitude higher than $CRB(\theta_o)$ and $CRB(\phi_o)$ in Fig. \ref{fig:CRB analysis}$\mathrm{b}$ and \ref{fig:CRB analysis}$\mathrm{c}$. Consequently, it is more difficult to estimate the ET orientation $\varphi$ compared with its directions ($\theta_o$, $\phi_o$), which is consistent with the conclusion we obtained in the 2D case \cite{Wang24TWC}.
\vspace{-10pt}
\subsection{Comparison of Beamforming Designs}
In this subsection, we compare the C\&S performance of our proposed beamforming designs and the other two baselines.

\begin{figure}[!t]
\centering
\includegraphics[width=3.5in]{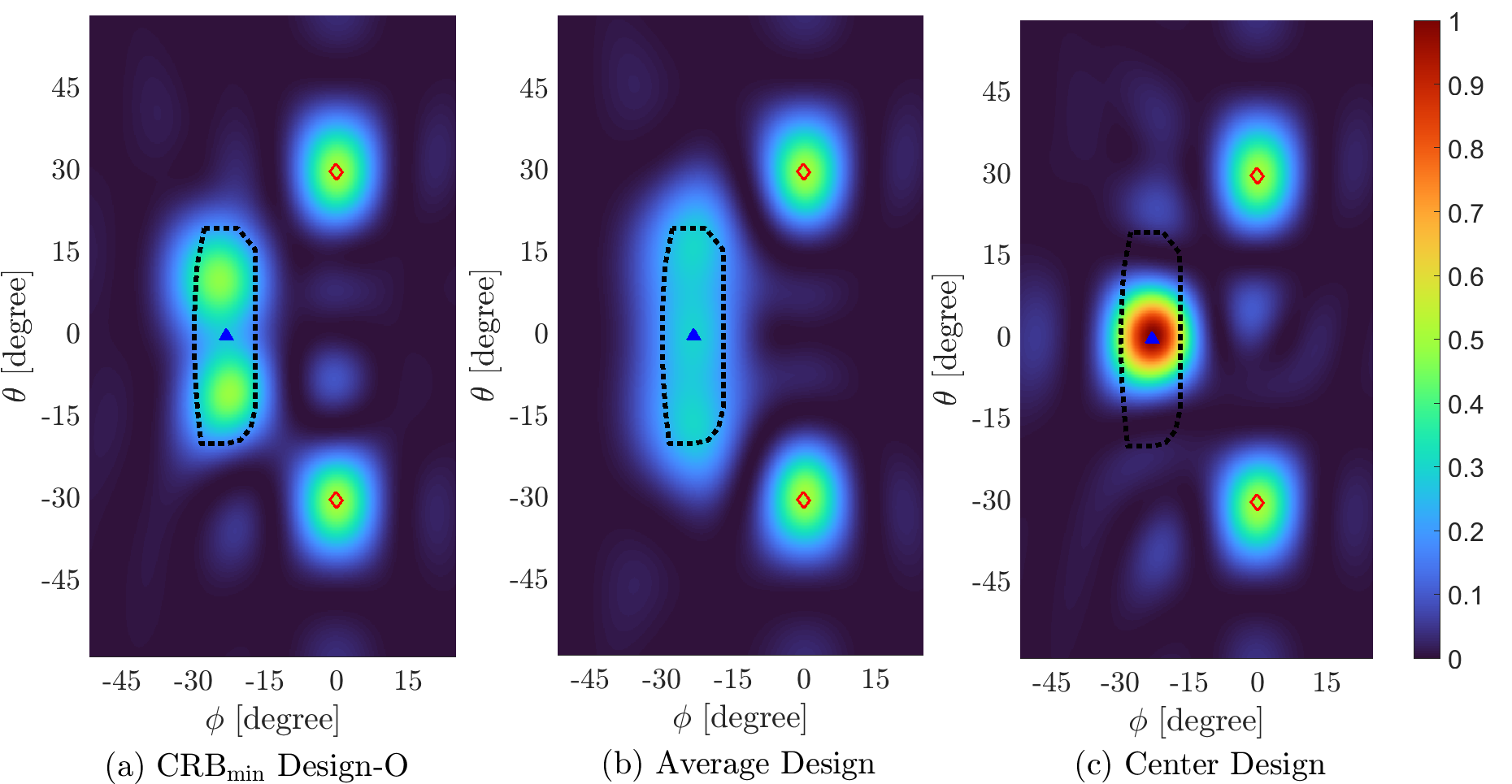}
\includegraphics[width=3.5in]{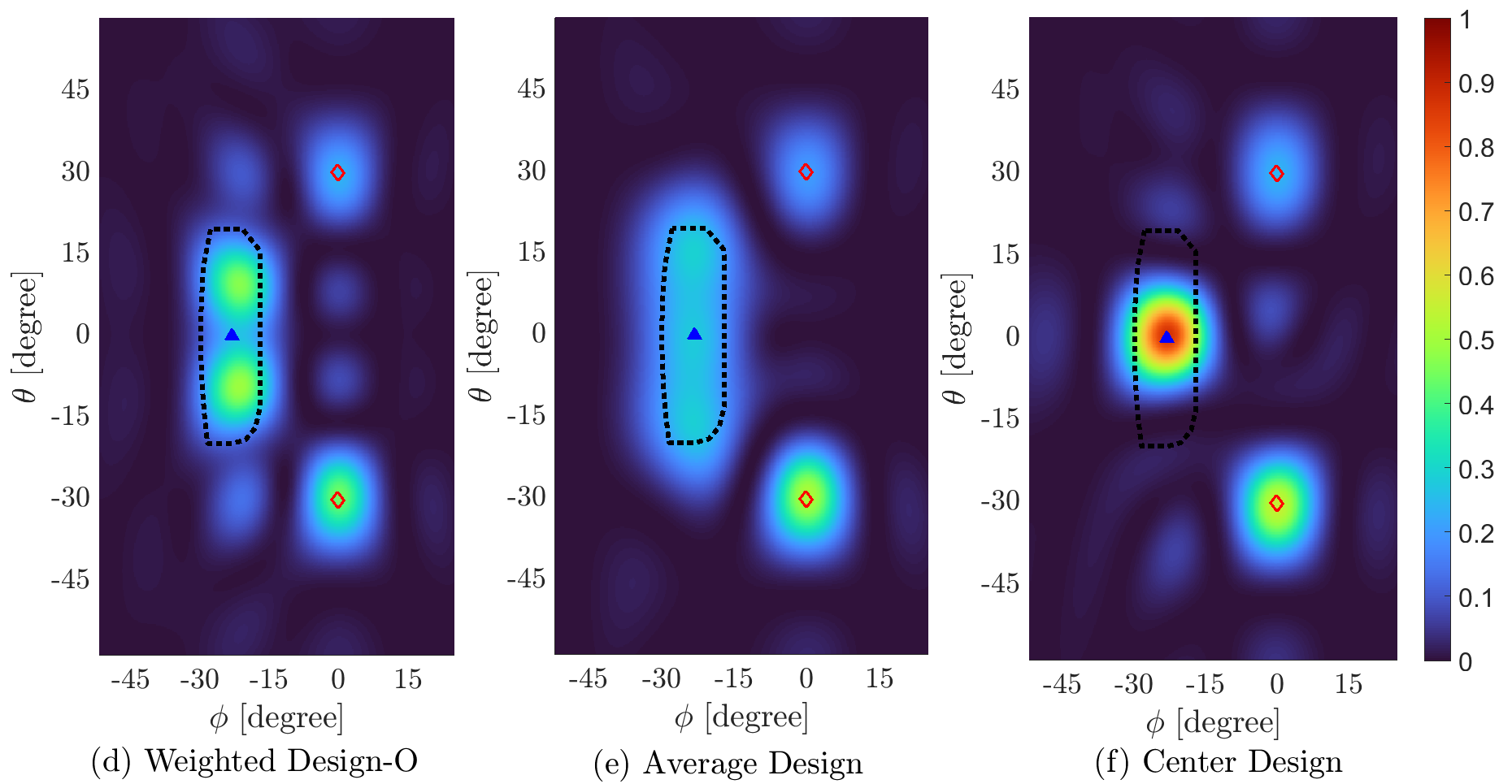}
\caption{Normalized beampatterns of different design methods of the vehicle-shaped ET for (a)-(c) The CRB-min problem $\mathcal{P}1.1$, the preset SINR threshold is $\Gamma=4\ \mathrm{dB}$; (d)-(e) The WIM problem $\mathcal{P}2.1$, the path loss for the CU at ($0^\circ,30^\circ$) is $3\ \mathrm{dB}$ greater than that of the CU at ($0^\circ,-30^\circ$). The red solid triangle and blue hollow rhombus refer to the directions of the ET center and CUs, respectively. The black dotted line defines the boundary of the visible ET surface.}
\label{fig:beampattern}
\end{figure}
\begin{figure}
\centering
\includegraphics[width=3.5in]{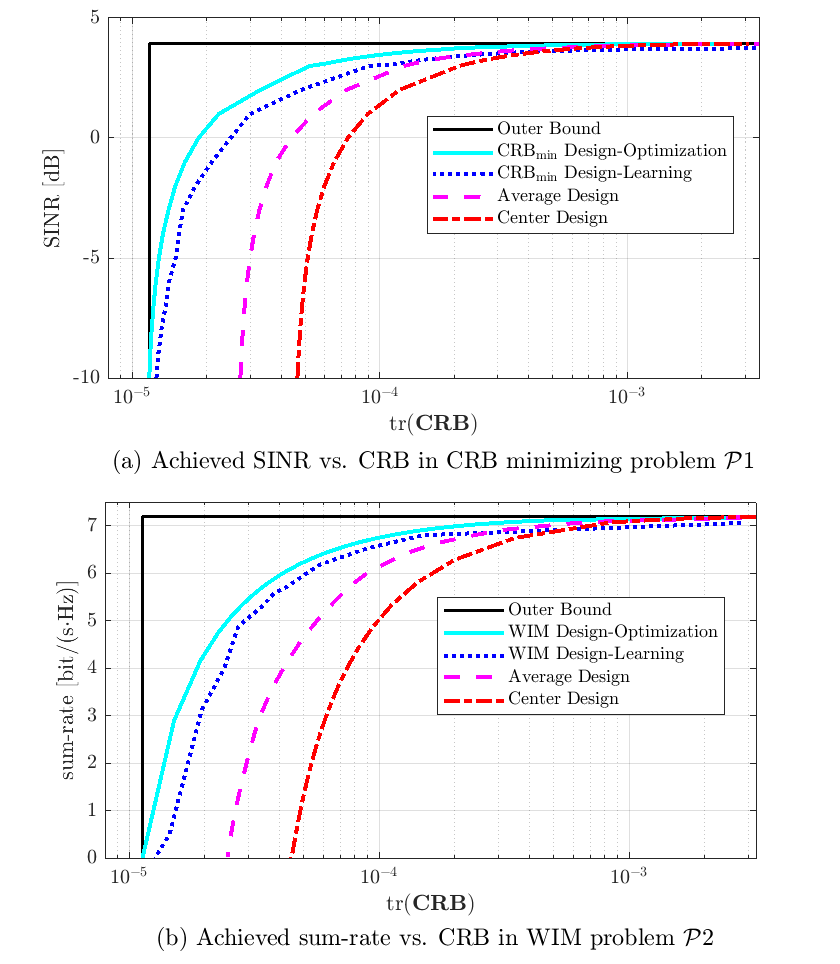}
\caption{C\&S performance trade-off.}
\label{fig:C_S_trade-off}
\end{figure}

Fig. \ref{fig:beampattern} illustrates the normalized 3D-beampatterns in the angular domains for three distinct beamforming designs implemented for the vehicle-shaped ET along with two CUs. It is observed that the mainlobe of $\mathrm{Center}$ $\mathrm{Design}$ solely points to the ET center regardless of the remaining areas, whereas the beam generated by $\mathrm{Average}$ $\mathrm{Design}$ covers an obviously larger area than the given ET boundary, leading to significant side lobes. In contrast, our proposed $\mathrm{CRB_{min}\ Design\mbox{-}O}$ creates two distinct main lobes that cover the front and rear regions of the ET, effectively suppressing energy leakage beyond the ET boundary. Although not shown here, the proposed $\mathrm{CRB_{min}\ Design\mbox{-}L}$ generates a sub-optimal beampattern for sensing the ET which resembles that of $\mathrm{CRB_{min}\ Design\mbox{-}O}$. For the communication functionality, solving $\mathcal{P}1$ results in a nearly equal resource distribution among all CUs, as shown in Figs.$\mathrm{\ref{fig:beampattern}a\mbox{-}\ref{fig:beampattern}c}$. In contrast, solving $\mathcal{P}2$ leads to a biased energy distribution, as illustrated in Figs.$\mathrm{\ref{fig:beampattern}d\mbox{-}\ref{fig:beampattern}f}$, where more resources are allocated to the CUs with better channel conditions.

Next, we investigate the C\&S performance trade-off of different beamforming designs in Fig. \ref{fig:C_S_trade-off}, where the outer bound is obtained by extending the maximal achievable SINR (sum rate) and minimal achievable CRB. For the CRB-min problem $\mathcal{P}1$ and WIM problem $\mathcal{P}2$, we use the achievable SINR threshold and the sum rate, respectively, to evaluate the communication capacity of different beamforming designs. We observe that the proposed $\mathrm{CRB_{min}\ Design\mbox{-}O}$ outperforms both baselines, significantly improving estimation accuracy while maintaining a desired level of communication quality. The confirms that, compared with the biased-centric beampattern of $\mathrm{CRB_{min}\ Design\mbox{-}O}$ in Fig. \ref{fig:beampattern}, the center-centric and average-coverage beampattern characteristics of $\mathrm{Center\ Design}$ and $\mathrm{Average\ Design}$ are not well-suited for estimating the kinematic parameters of ET. It should be noted that, the communication performance of $\mathrm{CRB_{min}\ Design\mbox{-}L}$ is slightly defective when approaching the maximal achievable SINR threshold and the maximal acheivable sum rate. Nevertheless, $\mathrm{CRB_{min}\ Design\mbox{-}L}$ still achieves near-optimal C\&S performance trade-off with only a minor gap between the C\&S curve of $\mathrm{CRB_{min}\ Design\mbox{-}O}$.

\begin{figure}[!t]
\centering
\includegraphics[width=3.5in]{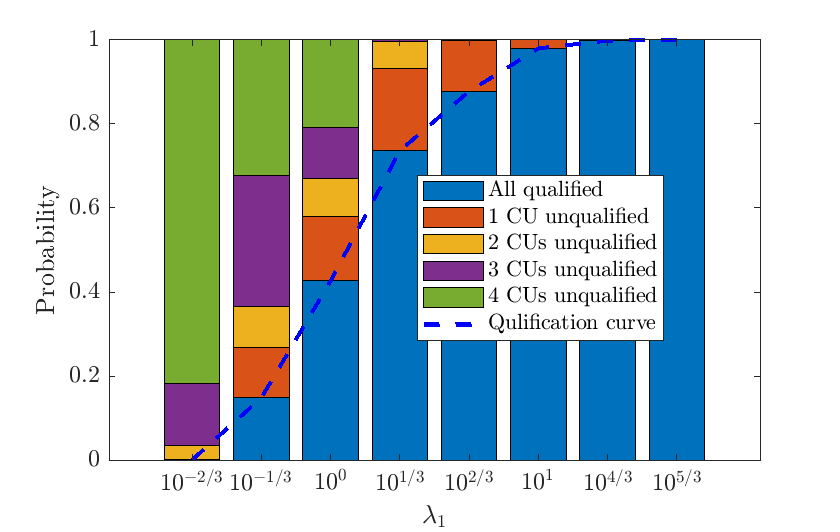}
\caption{The qualification probability of the SINR constraint $\mathrm{(\ref{SINR constraint})}$ versus the SINR penalty factor $\lambda_1$ in the CRB-min problem $\mathcal{P}1$.}
\label{fig:SINR probability}
\end{figure}
\begin{figure}[!t]
\centering
\includegraphics[width=3.5in]{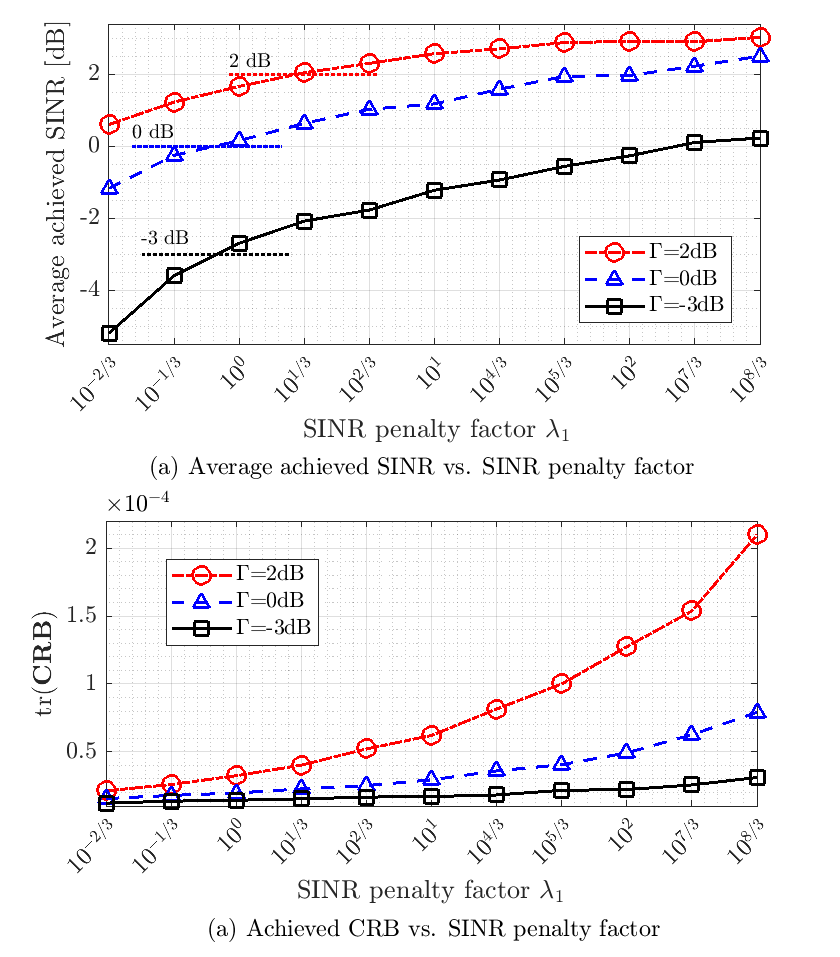}
\caption{C\&S performance versus SINR penalty factor $\lambda_1$. The ISACBeam-GNN is trained for CRB-min problem $\mathcal{P}1$.}
\label{fig:penalty}
\end{figure}

\begin{figure}[!t]
\centering
\includegraphics[width=3.5in]{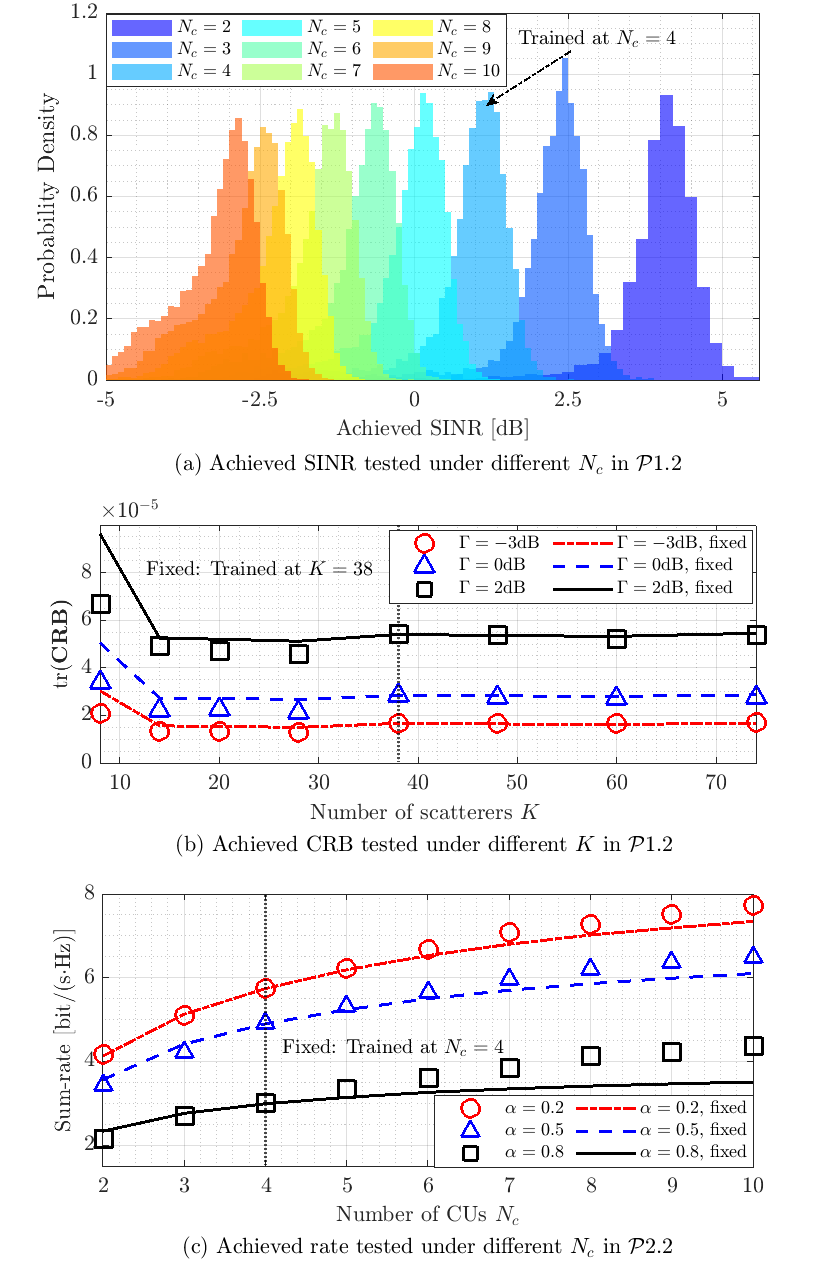}
\caption{Scalability of ISACBeam-GNN across varied numbers of CUs and scatterers. The fixed ISACBeam-GNN is trained under $\Gamma=0\ \rm{dB}$, $N_c=4$ and $K=38$ for $\mathcal{P}1$ and $\mathcal{P}2$. The non-fixed ISACBeam-GNN is trained and tested under identical conditions.}
\label{fig:generalization}
\end{figure}

\subsection{Performance of ISACBeam-GNN}
In this subsection, we assess the proposed ISACBeam-GNN through two aspects: the satisfaction of the SINR penalty term, and the scalability to varying numbers of CUs and scatterers.

Fig. \ref{fig:SINR probability} examines the probability of satisfying the minimum SINR constraint in $\mathcal{P}1$ under different penalty factors $\lambda_1$. The penalty factor for the beam coverage constraint is fixed at $\lambda_2=10$, ensuring that such constraint is nearly fully satisfied across our simulations. By increasing the value of $\lambda_1$, we notice that the probabilities of full and partial satisfaction are greatly improved. Specifically, when $\lambda_1$ reaches $10^{2/3}$, the occurrence of more than one CU failing to meet the minimum SINR disappears. Here, we adopt $\lambda_1=10$ throughout our simulation since an acceptable level of $98\%$ test samples can well meet the minimum SINR.

The value of the penalty factor $\lambda_1$ also has two-sided influence over the C\&S performance of the ISAC system, as is illustrated in Fig. \ref{fig:penalty}$\mathrm{a}$. To ensure full qualification of the minimum SINR constraint, ISACBeam-GNN prioritizes the CU with the worst channel conditions, which boosts the achieved SINR of other CUs beyond the SINR threshold at the same time. Such an effect is more pronounced when the SINR threshold is relatively low, for instance, at $\lambda_1=10^{8/3}$, the average achieved SINR is nearly twice the preset threshold of $\Gamma=-3\ \mathrm{dB}$. This is one major distinction between the optimization-based and the learning-based beamforming methods, where the SDR algorithm for 
$\mathcal{P}1$ precisely achieves the minimum SINR threshold at most cases. The misuse of wireless resources in ISACBeam-GNN inevitably leads to degraded sensing performance. As shown in Fig. \ref{fig:penalty}$\mathrm{b}$, the sensing CRB increases at larger $\lambda_1$, and such increase is more pronounced with larger SINR threshold, i.e., $\Gamma=2\ \mathrm{dB}$.

Last, we discuss the scalability of the proposed ISACBeam-GNN. For the CRB-min problem $\mathcal{P}1$ with $\Gamma = 0\ \mathrm{dB}$, Fig. \ref{fig:generalization}$\mathrm{a}$ plots the probability densities of achieved SINR across different numbers of CUs. We notice that the overall probability density histogram becomes wider and shifts towards higher SINR values as the testing number of CUs decreases. This indicates that, the optimized ISACBeam-GNN does not prioritize meeting the same SINR threshold when tested with varying CU counts. Instead, it focuses on maintaining a nearly constant total achievable communication rate across all CUs. Fig. \ref{fig:generalization}$\mathrm{b}$ shows that the well-trained ISACBeam-GNN achieves near-optimal performance when tested with a higher number of scatterers, whereas some minor performance degradation occurs when the number of scatterers is relatively smaller. Fig. \ref{fig:generalization}$\mathrm{c}$, on the other hand, illustrates that the testing performance of pre-trained ISACBeam-GNN for WIM problem $\mathcal{P}2$ remains nearly optimal only for a smaller number of CUs. As the number of CUs increases beyond the training configuration, the testing performance gradually deviates from the real value.

\section{Conclusion}
This paper considers a monostatic MIMO ISAC system, where one BS simultaneously communicates with multiple CUs and senses one 3D ET. We investigate the CRBs for estimating the kinematic parameters of the ET with an arbitrary but given shape, as well as the transmit beamforming design in the ISAC system. In particular, we develop an analytical second-order TFS model to describe the 3D ET surface, so as to quantitatively characterize the backscattering effects. Based on the surface model, we derive closed-form CRBs for estimating the ET kinematic parameters, including the center range, azimuth angle, elevation angle, and orientation. We present two beamforming design problems: the CRB-min problem and WIM problem. The first problem is tackled using the SDR technique, ensuring the minimization of sensing CRB subject to stringent constraints of the transmit power, communication SINR, and sensing beam coverage. The second problem, devoid of the SINR constraint, leverages the SCA strategy to minimize an objective combining the weighted communication sum rate and sensing CRB. To further reduce beamforming complexity, we propose a low-complexity learning approach, ISACBeam-GNN, with a novel separate-then-integrate architecture, where sensing and communication modules independently learn the structures of CRB and SINR (or sum rate), and the integration module balances the C\&S trade-off to generate the desired beamformers.

Through numerical simulation, we analyze the diverse CRB characteristics of vehicle-shaped and drone-shaped ETs, revealing their relations with PT. We also underscore the efficacy of our proposed beamforming designs, showcasing their ability to achieve superior C\&S trade-off and generate appropriate beampatterns for ET sensing. Notably, the learning-based ISACBeam-GNN offers a compelling and efficient alternative to optimization-based beamforming methods, demonstrating scalability across varying numbers of CUs and ET scatterers. Future work can consider specific estimation methods for ET parameters as well as dedicated signal design for sensing in ISAC systems.

{\appendices
\section*{Appendix \\ Proof of Proposition 1}
\subsection*{A.\hspace{5pt}General Derivation of EFIM $\mathbf{J}\left( {{\boldsymbol{\kappa }}} \right)$}
Define ${{\mathbf{q}}_{k}}=\mathbf{b}_k{{\mathbf{a}_k^H}}\mathbf{x}\left( t-{2{{d}_{k}}} / {c} \right)$, then the echo signal in $(\ref{echo})$ can be rewritten as $\mathbf{e}\left( t \right)=g\sum\nolimits_{k=1}^{K}{\sqrt{S_{k}}}{{\zeta }_{k}}{\mathbf{q}_{k}}$. With the definition of $\mathbf{F}_{\boldsymbol{\kappa}}=-\mathbb{E}\left[ \Delta _{{{\boldsymbol{\kappa }}}}^{{{\boldsymbol{\kappa }}}}\log p\left( {{\mathbf{y}}_{s}}|\boldsymbol{\xi } \right) \right]$, we have
\begin{align}
    \label{F_kappa}
   {{\mathbf{F}}_{{{\boldsymbol{\kappa }}}}}&=\frac{2{{g}^{2}}}{\sigma _{s}^{2}}\Re\left(\int_{{{t}_{s}}}{\sum\limits_{{k_{1}},{k_{2}}=1}^{K}{\mathbb{E}\left[ \zeta _{{{k}_{1}}}^{*}{{\zeta }_{{{k}_{2}}}} \right]\sqrt{{{S}_{{{k}_{1}}}}{{S}_{{{k}_{2}}}}}\frac{\partial \mathbf{q}_{{{k}_{1}}}^{H}}{\partial {{\boldsymbol{\kappa }}}}\frac{\partial {{\mathbf{q}}_{{{k}_{2}}}}}{\partial \boldsymbol{\kappa }^{T}}}}\text{d}t\right) \nonumber
\end{align}
\begin{align}
 & \stackrel{(a)}=\frac{2{{g}^{2}}}{\sigma _{s}^{2}}\sum\nolimits_{k=1}^{K}S_{k}\Re\left(\int_{{{t}_{s}}}{\frac{\partial \mathbf{q}_{k}^{H}}{\partial {{\boldsymbol{\kappa }}}}\frac{\partial {\mathbf{q}_{k}}}{\partial \boldsymbol{\kappa }^{T}}\text{d}t}\right),
\end{align}
where $(a)$ holds for $\mathbb{E}\left[ \zeta _{k}^{*}{\zeta }_{k} \right]=1$, $\mathbb{E}\left[ \zeta _{{{k}_{1}}}^{*}{{\zeta }_{{{k}_{2}}}} \right]=0$, $\forall {{k}_{1}}\ne {{k}_{2}}$. While ${{\mathbf{q}}_{k}}$ seems mathematically irrelevant with $\boldsymbol{\kappa}$, we utilize the chain rule $\partial\mathbf{q}_{k}^{H}/\partial{\boldsymbol{\kappa}}=\frac{\partial \mathbf{q}_{k}^{H}}{\partial {{\boldsymbol{\Theta }}_{k}}}\frac{\partial \boldsymbol{\Theta }_{k}^{T}}{\partial {{\boldsymbol{\kappa }}}}$, and instead switch to the calculation of ${\partial\boldsymbol{\Theta }_{k}^{T}}/{\partial {{\boldsymbol{\kappa }}}}$ in Appendix I-B and ${\partial\mathbf{q}_{k}^{H}}/{\partial {{\boldsymbol{\Theta }}_{k}}}$ in Appendix I-C. The intermediate variable is defined as ${\boldsymbol{\Theta}}_{k}=\left[ {{d}_{k}},{\theta }_{k},{\phi }_{k} \right]^T$, where ${{d}_{k}}$, ${{\theta }_{k}}$, ${{\phi}_{k}}$ are respectively the range, azimuth and elevation of the $k$-th scatterer. With Appendix I-B to Appendix I-D, we further rewrite $(\ref{F_kappa})$ as 
\begin{align}
\label{F_kappa_1}
&\nonumber\mathbf{F}_{\kappa}=\frac{2{{g}^{2}}}{\sigma _{s}^{2}}\sum\limits_{k=1}^{K}S_{k}\frac{\partial \boldsymbol{\Theta }_{k}^{T}}{\partial {{\boldsymbol{\kappa }}}}\Re\left(\int_{{t}_{s}}{\frac{\partial \mathbf{q}_{k}^{H}}{\partial {{\boldsymbol{\Theta }}_{k}}}\frac{\partial {{\mathbf{q}}_{k}}}{\partial \boldsymbol{\Theta }_{k}^{T}}\text{d}t} \right)\frac{\partial {{\boldsymbol{\Theta }}_{k}}}{\partial \boldsymbol{\kappa }^{T}}\\
&\hspace*{.6cm}=\frac{2{{g}^{2}}N_r}{\sigma _{s}^{2}}\sum\limits_{k=1}^{K}\Bigl[\nu_{1,k}\boldsymbol{\mu}_{1,k}\boldsymbol{\mu}_{1,k}^T+\nu_{2,k}\boldsymbol{\mu}_{2}\boldsymbol{\mu}_{2}^T\nonumber\\
&\hspace*{1cm}+\nu_{2,3,k}(\boldsymbol{\mu}_{2}\boldsymbol{\mu}_{3}^T+\boldsymbol{\mu}_{3}\boldsymbol{\mu}_{2}^T)+\nu_{3,k}\boldsymbol{\mu}_{3}\boldsymbol{\mu}_{3}^T\Bigr].
\end{align}

Similar to the steps in $(\ref{F_kappa})$, $\mathbf{f}_{\bm{\kappa},g}$ and ${{f}_{g}}$ can be obtained as
\begin{align}
&\mathbf{f}_{\bm{\kappa},g}=\frac{2g}{\sigma_s^2}\sum_{k=1}^KS_k\frac{\partial\bm{\Theta}_k^H}{\partial\bm{\kappa}}\Re\int_{t_s}\frac{\partial\mathbf{q}_k^H}{\partial \mathbf{\Theta}_{k}}\mathbf{q}_kdt=\frac{2gt_sN_r}{\sigma_s^2}\boldsymbol{\mu}_4,\\
\label{f_g}
&f_g=\frac{2t_s}{\sigma_s^2}\sum_{k=1}^KS_k\mathbf{q}_k^H\mathbf{q}_k=\frac{2t_sN_r}{\sigma_s^2}\nu_4^{-1}.
\end{align}

Combining $(\ref{F_kappa_1})-(\ref{f_g})$ with $(\ref{CRB})$, we obtain EFIM in $(\ref{FIM})$.

\subsection*{B.\hspace{5pt}Derivation of ${\partial{\boldsymbol{\Theta} }_{k}^{T}}/{\partial \boldsymbol{\kappa}}$}
Decompose $\partial \boldsymbol{\Theta }_{k}^{T}/\partial {{\boldsymbol{\kappa }}}=\left[ \partial {{d}_{k}}/\partial {{\boldsymbol{\kappa }}},\partial {{\theta }_{k}}/\partial {{\boldsymbol{\kappa }}},\partial {{\phi }_{k}}/\partial {{\boldsymbol{\kappa }}} \right]$ as
\begin{align}
\frac{\partial{{{d}}_{k}}}{\partial {{\boldsymbol{\kappa }}}}&=\left(\frac{\partial{d_k^2}}{\partial{d_k}}\right)^{-1}\frac{\partial{d_k^2}}{\partial\boldsymbol{\kappa}}\approx \boldsymbol{\mu}_{1,k},\\
\frac{\partial {{\theta }_{k}}}{\partial {{\boldsymbol{\kappa }}}}&=\frac{\partial \arctan\left(p_{y,k}/p_{x,k}\right)}{\partial\left(p_{y,k}/p_{x,k}\right)}  \frac{\partial\left(p_{y,k}/p_{x,k}\right)}{\partial\boldsymbol{\kappa}}\approx \boldsymbol{\mu}_2,\\
\frac{\partial {{\phi }_{k}}}{\partial {{\boldsymbol{\kappa }}}}&=\frac{\partial\arcsin{\left(p_{z,k}/d_k\right)}}{\partial{\left({p_{z,k}}/{d_k}\right)}}\frac{\partial{\left({p_{z,k}}/{d_k}\right)}}{\partial\bm{\kappa}}\approx \boldsymbol{\mu}_3,
\end{align}
where $\boldsymbol{\rho}_k$, $p_{x,k}$ are the local position and the $x$ coordinate in the global frame for the $k$-th scatterer, ${d}_{\bot,k}=(p_{x,k}^2+p_{y,k}^2)^{1/2}$ is the ET range within the global $Oxy$ plane. The approximations are valid under the assumption that $|p_{z,k}| \ll d_{\bot,k}$, $d_{o}\approx d_{k}$, $\forall k$ and ${1}/{d_o} \rightarrow 0$.

\subsection*{C.\hspace{5pt}Derivation of $\Re\left(\int_{{{t}_{s}}}{\frac{\partial \mathbf{q}_{k}^{H}}{\partial {{\boldsymbol{\Theta }}_{k}}}\frac{\partial {{\mathbf{q}}_{k}}}{\partial \boldsymbol{\Theta }_{k}^{T}}\mathrm{d}t}\right)$ and $\Re\left(\int_{{{t}_{s}}}{\frac{\partial \mathbf{q}_{k}^{H}}{\partial {{\boldsymbol{\Theta }}_{k}}}{{\mathbf{q}}_{k}}\mathrm{d}t}\right)$}
Start with the calculation of $\frac{\partial {{\mathbf{q}}_{k}}}{\partial \boldsymbol{\Theta }_{k}^{T}}=\left[ \frac{\partial {{\mathbf{q}}_{k}}}{\partial {{d}_{k}}},\frac{\partial {{\mathbf{q}}_{k}}}{\partial {{\theta }_{k}}},\frac{\partial {{\mathbf{q}}_{k}}}{\partial {{\phi }_{k}}} \right]$
\begin{align}
\frac{\partial {{\mathbf{q}}_{k}}}{\partial {{d}_{k}}}&=-\frac{2}{c}{{\mathbf{b}}_{k}}\mathbf{a}_{k}^{H}\mathbf{\dot{x}}\left( t-\frac{2{{d}_{k}}}{c} \right),\\
\frac{\partial {{\mathbf{q}}_{k}}}{\partial {\theta_k\vert\phi_k}}&=\left[ \left({\dot{\hat{\mathbf{b}}}}_{k}|{\dot{\bar{\mathbf{b}}}}_{k}\right)\mathbf{a}_{k}^{H}+{{\mathbf{b}}_{k}}\left(\dot{\hat{\mathbf{a}}}_{k}^{H}|\dot{\bar{\mathbf{a}}}_{k}^{H}\right) \right]\mathbf{x}\left( t-\frac{2{{d}_{k}}}{c} \right),
\end{align}
where $\mathbf{\dot{x}}\left( t \right)=\partial \mathbf{x}\left( t \right)/\partial t$. With the identities in Appendix I-D, $\Re\left(\int_{{{t}_{s}}}{\frac{\partial \mathbf{q}_{k}^{H}}{\partial {{\boldsymbol{\Theta }}_{k}}}\frac{\partial {{\mathbf{q}}_{k}}}{\partial \boldsymbol{\Theta }_{k}^{T}}\text{d}t}\right)$ can be formulated as
\begin{align}
&\Re\left(\int_{{{t}_{s}}}{\frac{\partial \mathbf{q}_{k}^{H}}{\partial {{d}_{k}}}\frac{\partial {{\mathbf{q}}_{k}}}{\partial {{d}_{k}}}\text{d}t}\right)=N_r S_k^{-1}\nu_{1,k},\\
&\Re\left(\int_{{{t}_{s}}}{\frac{\partial \mathbf{q}_{k}^{H}}{\partial {{\theta }_{k}}}\frac{\partial {{\mathbf{q}}_{k}}}{\partial {{\theta }_{k}}}\text{d}t}\right)=N_r S_k^{-1}\nu_{2,k},\\
&\Re\left(\int_{{{t}_{s}}}{\frac{\partial \mathbf{q}_{k}^{H}}{\partial {{\phi }_{k}}}\frac{\partial {{\mathbf{q}}_{k}}}{\partial {{\phi }_{k}}}\text{d}t}\right)=N_r S_k^{-1}\nu_{3,k},\\
&\Re\left(\int_{{{t}_{s}}}{\frac{\partial \mathbf{q}_{k}^{H}}{\partial {{d }_{k}}}\frac{\partial {{\mathbf{q}}_{k}}}{\partial {\theta}_{k}|{\phi}_{k}}\text{d}t}\right)=0,\\
&\Re\left(\int_{{{t}_{s}}}{\frac{\partial \mathbf{q}_{k}^{H}}{\partial {{\phi }_{k}}}\frac{\partial {{\mathbf{q}}_{k}}}{\partial {{\theta }_{k}}}\text{d}t}\right) =N_r S_k^{-1}\nu_{2,3,k}.
\end{align}

Similar, we derive $\Re\left(\int_{{{t}_{s}}}{\frac{\partial \mathbf{q}_{k}^{H}}{\partial {{\boldsymbol{\Theta }}_{k}}}{{\mathbf{q}}_{k}}\mathrm{d}t}\right)$ as
\begin{align}
&\Re\left(\int_{{{t}_{s}}}{\frac{\partial \mathbf{q}_{k}^{H}}{\partial {{d}_{k}}}{{\mathbf{q}}_{k}}\text{d}t}\right)=0,\\
&\Re\left(\int_{{{t}_{s}}}{\frac{\partial \mathbf{q}_{k}^{H}}{\partial {{\theta}_{k}}|{{\phi}_{k}}}{{\mathbf{q}}_{k}}\text{d}t}\right)={{N}_{r}t_s}\Re\left( \left(\dot{\hat{\mathbf{a}}}_{k}|\dot{\bar{\mathbf{a}}}_{k}^{H}\right)\mathbf{R}_x\mathbf{a}_{k}\right).
\end{align}

\subsection*{D.\hspace{5pt}Related Identities}
To obtain the tightest bound, the center of the UPA is set as the reference element with zero phase. Then, the transmit antenna response can be presented as
\begin{align}
    &\mathbf{a}_{k}=\mathbf{a}_{k,z}\otimes\mathbf{a}_{k,x}^T,\\
    &\mathbf{a}_{k,x}=\exp \left( j\pi \left( {{N}_{t,x}}-1 \right)\mathcal{F}^{1,0}_k/2 \right) \nonumber\\
    &\hspace*{.2cm}{{\left[ 1,\exp\left( -j\pi \mathcal{F}^{1,0}_k \right),...,\exp\left( -j\pi\left( {{N}_{t}}-1 \right) \mathcal{F}^{1,0}_k \right) \right]}^{T}},\\
    &\mathbf{a}_{k,z}=\exp \left( j\pi \left( {{N}_{t,z}}-1 \right)\sin\phi_k/2 \right) \nonumber\\
    &\hspace*{.0cm}{{\left[ 1,\exp\left( -j\pi \sin {{\phi }_{k}} \right),...,\exp\left( -j\pi\left( {{N}_{t}}-1 \right) \sin {{\phi }_{k}} \right) \right]}^{T}}.
\end{align}

Next, with the assumption of $\mathbb{E} \left[\mathbf{s}(t) \mathbf{s}^H (t) \right] = \mathbf{I}_{N_c}$, we obtain $\int_{t_s} \mathbf{s}(t) \mathbf{s}^H (t) \text{d}t = t_s\mathbf{I}_{N_c}$. With a sufficiently long observation period $t_s$, we have the Fourier transform \cite{Wang24TWC} as
\begin{align}
  & \int_{t_s}{s_{{c}_{1}}^{*}\left( t-\tau  \right){{{\dot{s}}}_{{{c}_{2}}}}\left( t-\tau  \right)\text{d}t} \approx 0, \forall n_{1,2},\\
&\int_{t_s}{\dot{s}_{{c}_{1}}^{*}\left( t-\tau  \right){\dot{s}_{{c}_{2}}}\left( t-\tau  \right)\text{d}t} \approx\left\{
\begin{aligned}
&{\left( 2\pi B \right)}^{2}, && {c}_{1}={c}_{2,}  \\
&0, && {c}_{1}\ne {c}_{2,}
\end{aligned}
\right.
\forall n_{1,2,}
\end{align}
where $\tau$ is the time delay.
\bibliographystyle{IEEEtran}
\bibliography{Reference1.bib}
\end{document}